\documentclass{jfm}
\usepackage{lineno}
\usepackage{graphicx}
\usepackage[update,prepend]{epstopdf}
\usepackage{epstopdf,epsfig}
\usepackage{xcolor} 
\usepackage{enumitem}
\usepackage{amsmath}
\usepackage{amsmath,bm} 
\usepackage{multirow} 
\usepackage{booktabs} 
\usepackage{amsbsy} 
\usepackage{mathtools} 
\usepackage{amssymb} 
\usepackage{natbib} 
\usepackage{amsfonts} 
\usepackage{amsgen} 
\usepackage{bookmark}
\usepackage{textgreek}
\usepackage{hyperref}
\graphicspath{{figure/}} 
\hypersetup{colorlinks, linkcolor=red, citecolor=blue, urlcolor=green}

\shorttitle{Optimal growth}
\shortauthor{X. Chen, B. Wan, G. Tu, M. Duan, X. Li, J. Chen}

\title{Non-modal growth analysis of high-speed flows over an inclined cone}

\author{Xi Chen \and Bingbing Wan\corresp{\email{wanbing@tju.edu.cn}} \and Guohua Tu \and Maochang Duan \and Xiaohu Li \and \and Jianqiang Chen\corresp{\email{chenjq@cardc.cn}}}

\affiliation{\aff{1}State Key Laboratory of Aerodynamics, Mianyang, Sichuan 621000, China}
\begin{document}

\maketitle

\begin{abstract}
Spatial optimal responses to both inlet disturbances and harmonic external forcing for hypersonic flows over a blunt cone at nonzero angles of attack are obtained by efficiently solving the direct-adjoint equations with a parabolic approach. In either case, the most amplified disturbances initially take the form of localized streamwise vortices on the windward side and will undergo a two-stage evolution process when propagating downstream: they first experience a substantial algebraic growth by exploiting the Orr and lift-up mechanisms, and then smoothly transition to a quasi exponential-growth stage driven by the crossflow-instability mechanism, accompanied by an azimuthal advection of the disturbance structure towards the leeward side. The algebraic-growth phase is most receptive to the external forcing, whereas the exponential-growth stage relies on the disturbance frequency and can be significantly strengthened by increasing the angle of attack. The wavemaker delineating the structural sensitivity region for the optimal gain is shown to lie on the windward side immediately downstream of the inlet, implying a potent control strategy. Additionally, considerable non-modal growth is also observed for broadband high-frequency disturbances residing in the entropy layer.
\end{abstract}


\section{Introduction}
Hypersonic vehicles are faced by three-dimensional (3-D) boundary layer transition problems that yield an undesirable increase in heat flux and skin friction. In contrast to the two-dimensional (2-D) boundary layer over a flat plate, cone at zero degrees angle of attack, etc, the 3-D counterpart features nonuniform pressure distribution in the azimuthal direction, and as such exhibits three distinct regions across the vehicle surfaces, i.e., the attachment-line region (high-pressure region), the vortex region (low-pressure region) and the crossflow region in between. Each region sustains certain modal instabilities likely responsible for the boundary-layer breakdown. Of particular interest is the crossflow region where boundary-layer transition is generally dominated by crossflow instabilities comprising stationary and traveling modes \citep{Reed1989} and occurs over the largest part of the vehicle surface.
\par
Base flow in the crossflow region varies mildly in the azimuthal direction, which appears amenable to one-dimensional (1-D) stability analyses. To obtain the evolution of crossflow vortices, it is common practice to first model a vortex path and the azimuthal wavenumber variation along the path, and then integrate the (nonlinear) parabolized stability equations (PSE) assuming azimuthal periodicity \citep{Kocian2019,Song2020}. However, the 1-D tools find it difficult to model the typical wavepacket structures of the crossflow mode, and also could not deal with the nonlinear interactions between multiple modes as local modes may have different trajectories. Therefore, one attempts to establish a global stability analysis framework to delineate the crossflow instability behaviour in a 3-D boundary layer. \cite{Paredes2016a} and later \cite{Lakebrink2017} adopted BiGlobal approach solving 2-D eigenvalue equations to uncover the global modal structure of the traveling crossflow instability over an elliptic cone. They found that the shape function of the crossflow mode exhibits rapid oscillations in the azimuthal direction and covers almost the entire surface of the vehicle, thus requiring a large number of grid points to resolve. In a similar study on a lifting body, \cite{Chen2022} documented that the global crossflow modal eigenvalues are too sensitive to azimuthal resolution to be grid converged. This extreme sensitivity of eigenvalues reflects, on the one hand, the highly non-orthogonal nature of the linear operator and, on the other hand, is physically due to the weakly non-parallel effects (i.e., the azimuthal wavelength of the crossflow mode is much smaller compared to the azimuthal variation scale of the base flow)\citep{Chomaz2005}, analogous to previous studies on streamwise BiGlobal problem \citep[for instance]{Akervik2008}. They also showed that the global spectrum for a single frequency contains many unstable modes without a predominant one, which was reported by \cite{Liu2022} as well in studying an inclined cone configuration. The poor grid convergence of eigenvalues and the multiplicity of crossflow modes make the interpretation of global modal instability results challenging. As a result, the global modal stability analysis may be unable to properly describe the crossflow instability in a complex 3-D boundary layer. Furthermore, it has become increasingly evident that modal instability analysis alone can not explain all transition phenomena, as exemplified by the famous blunt-cone paradox concerning subcritical transition on a large-bluntness cone \citep{Reshotko2000}. Recently, \cite{Paredes2023JSR} and \cite{Chen2023b} who investigated the leeward modal instabilities of hypersonic flow over yawed blunt cones, both identified strikingly low $N$-factors (below 2) correlated with the observed transition location, pointing to unknown transition mechanisms.
\par
The alternative approach is based on the non-modal stability analysis framework which considers the interferences of multiple modes simultaneously \citep{Schmid2001,Schmid2007}. Instead of focusing on the modal instability, the non-modal stability analysis aims to identify the most dangerous case from all potential routes of disturbance evolution, which is important since inlet conditions are typically unavailable in realistic situations. Moreover, the optimal solution proves to be robust to small perturbations even if the eigenvalues may not \citep{Reddy1993,Schmid2007}. The non-modal stability analysis comprises two main branches, i.e., the optimal growth analysis and the input-output/resolvent analysis (abbreviated as I-O analysis). The optimal growth analysis finds the intrinsic disturbances in the flow field that result in the maximal energy growth over a certain time interval (temporal approach) or distance downstream (spatial approach). On the other hand, the I-O analysis \citep{Jovanovic2005} focuses on determining the optimal external forcing (input) that leads to the largest response (output, measured by disturbance energy) in the flow field. Albeit with different physical meanings and focuses, both branches share a similar mathematical framework \citep{Schmid2007} and even yield the same results when restricting the input and output at the inlet and outlet, respectively \citep{Bugeat2019}. Hereafter we will concentrate our attention on the optimal growth analysis.
\par
The early non-modal stability analyses were mostly conducted for simple shear flows (Poiseuille flow, Couette flow, Blasius boundary layer, etc) in the low-speed regime, showing that disturbances can be considerably (linearly) amplified through the Orr mechanism and lift-up effect even if they are modally stable \citep{Schmid2001}. The Orr mechanism is predominant for 2-D waves and relies on the deformation of vorticity patches under the action of the base flow. The deformation reduces the support area of the vorticity such that the vorticity extrema increase owing to the Kelvin circulation preservation theorem, thereby causing a transient energy growth \citep{Nastro2020}. The lift-up effect corresponds to the emergence of streaks induced by streamwise vortices. In general, 3-D optimal perturbation combines these mechanisms in a synergistic manner in which the cross-stream velocity produced by the Orr mechanism enhances the streamwise rolls associated with streak production \citep{Farrell1993}. The optimal perturbation can sometimes initiate the modal instability, especially in 3-D boundary layers where modal and non-modal growths are observed for similar disturbance structures. \cite{Corbett2001} who studied temporal optimal disturbances in the Falkner-Skan-Cooke flow modeling the boundary layer over an infinite swept wing found that these initially take the form of vortices which are almost aligned with the external streamline and evolve into crossflow modes downstream. They argued that non-modal growth may provide proper initial conditions for modal growth and thus constitutes a preferential receptivity path for the selection of exponential instabilities in 3-D boundary layers. On a similar configuration, \cite{Tempelmann2010} utilized a spatial approach based on the method of Lagrange multipliers and PSE to reveal that the optimal disturbance manifests as streamwise vortices aligned against the crossflow, extracts energy by leveraging both the Orr and lift-up mechanisms, and develops smoothly into crossflow modes downstream. They \citep{Tempelmann2012} later extended the study to compressible flow (up to Mach 1.5), showing that optimal growth increases for higher Mach numbers and wall cooling destabilizes disturbances of non-modal nature. Nonetheless, the physical mechanisms of optimal growth in compressible boundary layers are found to be similar to the incompressible counterpart.
\par
The first non-modal stability analysis in compressible flow comes from \cite{Hanifi1996} who, adopting the temporal approach, examined the effects of different parts of spectra on the optimal growth, and obtained similar optimal structures as the incompressible flow. Their work was later complemented by a spatial one performed by \cite{Tumin2003}. \cite{Laible2016} identified a constantly non-modal growth mechanism in the streamwise development of the streak mode in oblique breakdown, by which the streak mode retains a higher amplitude level than other nonlinearly generated modes. \cite{Paredes2019,Paredes2020} observed substantial non-modal growth of disturbances peaking within the entropy layer around the nose of a blunt cone, and furthermore showed that such disturbances could trigger the boundary layer transition through oblique breakdown.
\par

   The aforementioned studies considered essentially 2-D compressible boundary layers where the flow field is treated as homogeneous in at least one direction (usually the azimuthal/spanwise direction). Non-modal analyses on truly 3-D compressible boundary layers are rare. \cite{Cook2020} obtained the optimal response of a hypersonic finned-cone boundary layer to wall roughnesses by partitioning the entire domain into smaller segments and progressively solving the I-O equations of each segment. \cite{Paredes2022} calculated the optimal disturbances for flow field around an isolated 3-D roughness with harmonic linearized Navier-Stokes (HLNS) equations, showing that the optimal disturbances amplify most rapidly in the backward-flow region of the roughness. The above two studies both entail iteratively solving large-scale linear systems with unknowns at order of one million, which are prohibitively expensive especially for a parametric study. \cite{Quintanilha2022} performed temporal non-modal stability analysis on an elliptic-cone boundary layer under hypersonic flight conditions. They observed a strong transient growth in the crossflow region, which is strengthened with increasing the flight altitude and as such is likely the important mechanism responsible for the high-altitude boundary-layer transition. In their study, the optimal disturbance is constructed by a linear combination of a finite number of global modes whose amplitude coefficients are optimized to achieve the maximum amplification at a given time horizon. Their results may, however, be compromised by the aforementioned high sensitivity of global crossflow modes, as well as the partial consideration of the eigenvalue spectrum \citep{Bagheri2009} since obtaining the entire spectrum is infeasible. Moreover, the spatial optimal growth pertaining to crossflow disturbances in a truly 3-D boundary layer remains largely unknown.
   \par
   In the present study, we aim at establishing a robust description of the crossflow instabilities from a global perspective with help of the non-modal analysis, and furthermore, uncovering the optimal growth mechanisms for a prototypical 3-D boundary layer. To this end, we attempt to constitute an efficient yet accurate spatial optimal growth calculation framework based on a parabolic set of equations, and employ it to systematically study the optimal responses over a hypersonic inclined cone to both inlet perturbation and external forcings, with a special focus on the influence of the angle of attack (AoA). The related methodologies are summarized in \S\ref{sec:2}. Then, the linear non-modal characteristics of the selected configuration are presented in \S\ref{sec:3}, followed by discussions on the effects of AoA in \S\ref{sec:4}. Finally, concluding remarks and discussions are given in \S\ref{sec:5}.
\section{Numerical set-up and linear stability theories}\label{sec:2}
\subsection{Model}
\begin{figure}
\centering
\includegraphics[width = 0.95\textwidth]{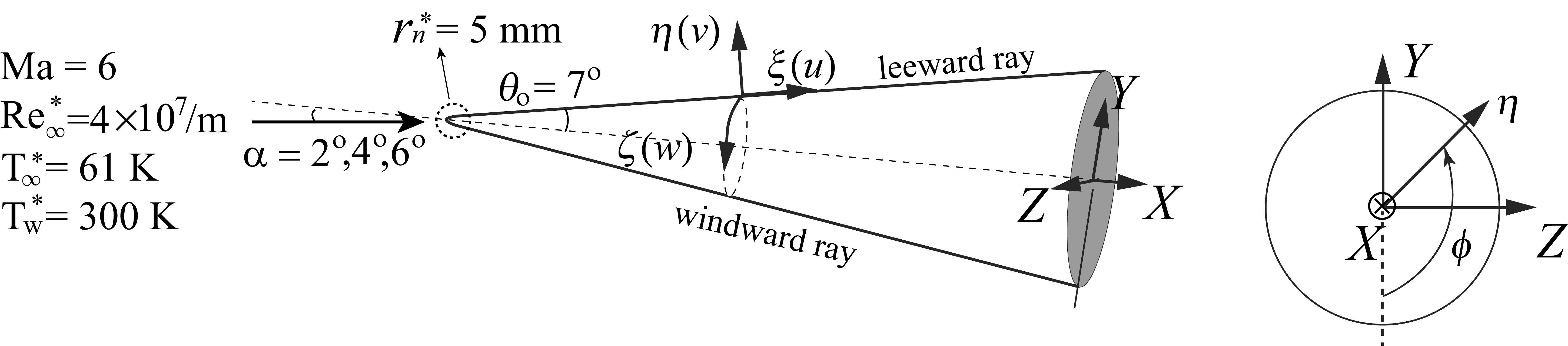}
\caption{Sketch of the flow configuration. The velocity components ($u,v,w$) in the body-oriented coordinates ($\xi,\eta,\zeta$) are also shown. } 
\label{model}
\end{figure}
The model currently being studied is a 7$^\circ$-half-angle blunt cone at angles of attack of 2$^\circ$, 4$^\circ$ and 6$^\circ$, respectively. The model is 900 mm long and has a nose radius of 5 mm. The freestream Mach number $Ma = 6$, unit Reynolds number $Re_\infty^* = 4\times10^7$/m, freestream static temperature $T_\infty^* = 61.1$ K, the wall temperature $T_w^*$ = 300 K. The detailed flow configuration can be found in figure \ref{model}. The dimensional variables are denoted with the upscript $()^*$. All flow quantities are non-dimensionalized using the reference length $L^*$ = 0.0387 mm (the boundary layer scale at $X^* = 10$ mm), the freestream velocity, density and temperature.
\subsection{Base flow calculations}
A well-validated in-house code \citep{Zhao2017} was used to solve the unsteady compressible Navier-Stokes equations. The grid resolution is $N_\xi\times N_\eta\times N_\zeta = 501\times301\times301$ for the first two cases and $N_\xi\times N_\eta\times N_\zeta = 751\times301\times361$ for the last case (6 deg AoA), where $N_\xi,N_\eta,N_\zeta$ is the grid number in the streamwise, wall-normal and azimuthal direction, respectively. The grid is stretched in the wall-normal and streamwise directions so that sufficient (at least 100) points are in the boundary layer and in the vicinity of the nose part, and is furthermore aligned with the leading-edge shock to diminish the numerical oscillations from the shock layer. Only half part of the flow field is simulated by exploiting the azimuthal symmetry. The inviscid fluxes are computed by using a third-order WENO scheme, while the viscous fluxes are discretized using a fourth-order central-difference scheme. The time integration is performed using a
third-order Runge-Kutta scheme. The grid convergence has been tested by increasing the grid points along the azimuthal and wall-normal directions.
 \par


\begin{figure}
\centering
\includegraphics[width = 0.99\textwidth]{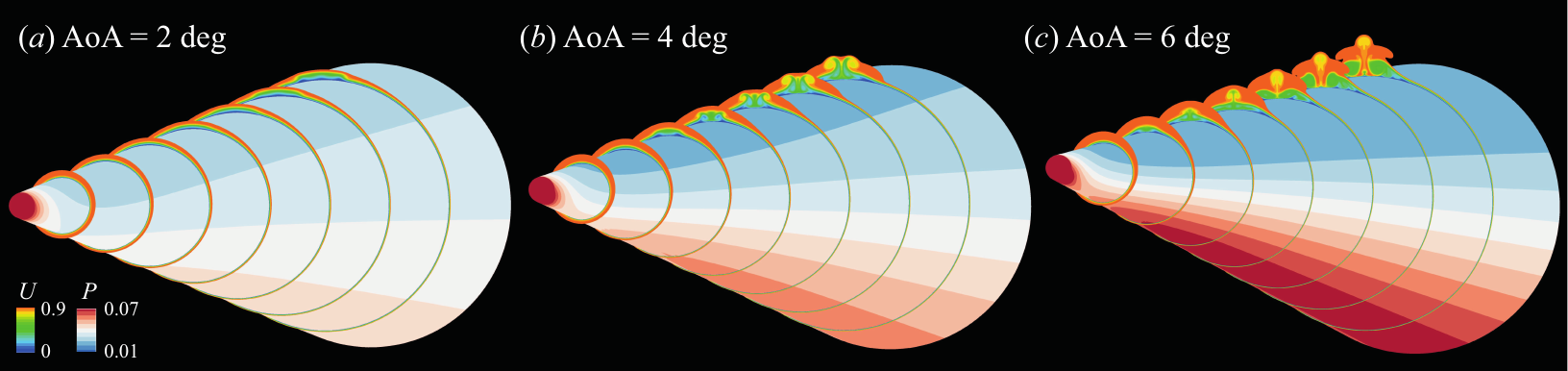}
\caption{Contours of the wall pressure and axial velocity at several axial locations from $X^* = 50$ mm to $X^* = 350$ mm for three cases with different angles of attack: ($a$) 2 deg, ($b$) 4 deg, ($c$) 6 deg. } 
\label{flowfield}
\end{figure}
The flow fields are illustrated by the surface pressure distribution and the axial velocity distribution at several cross sections in figure \ref{flowfield}. As expected, the azimuthal pressure gradient is present, driving the fluid towards the leeward region. Increasing the angle of attack strengthens the pressure gradient, leading to a more prominent vortical structure in the vicinity of the leeward ray.
\subsection{Governing equations of disturbances}
We concentrate on the linearized compressible Navier-Stokes equations describing the evolution of infinitesimal perturbations in a 3-D geometry with two rapidly varying (wall-normal and azimuthal) directions and one slowly varying (streamwise) direction. Decompose the flow state $\boldsymbol {Q}
\equiv (\rho,u,v,w,T)^{tr}$ into the base flow $\boldsymbol {\bar q} \equiv (\bar\rho,\bar u,\bar v,\bar w,\bar T)^{tr}$ plus the infinitesimal disturbances $\boldsymbol {q'} = (\rho', u', v', w', T')^{tr}$, where $tr$ stands for the transpose and $u,v,w$ describe the velocity components in the streamwise, wall-normal and azimuthal direction, respectively. The set of governing equations is then conveniently written in the form
\begin{equation}\label{LNS}
\frac{\partial \boldsymbol{q'}}{\partial t} = \mathcal{ {L}}\boldsymbol q' + \boldsymbol f,
\end{equation}
where $\mathcal{L}$ is the linearized Navier-Stokes operator; $\boldsymbol f$ denotes the external forcing.
\par
Our intention is to find a set of linear governing equations that is parabolic in nature. Such a system can be solved efficiently by employing marching techniques, and allows for extensive parametric studies. To this end, we can take the PSE ansatz since the base flow varies slowly in the streamwise direction
\begin{equation}\label{ansatz}
\boldsymbol q'(\xi,\eta,\zeta,t) = \boldsymbol {\hat q}(\xi,\eta,\zeta)\exp(i\int_{\xi_0}^{\xi}\alpha(s){\rm{d}}s-i\omega t) + c.c.,
\end{equation}
where $c.c.$ designates the complex conjugate part. The real, streamwise varying wavenumber $\alpha(\xi)$ is determined iteratively by
\begin{equation}\label{alpha}
\alpha_{new} = \alpha_{old} + {\rm{Im}}(\frac{1}{E}\frac{\partial E}{\partial\xi}),
\end{equation}
 with $E$ denoting the total disturbance energy defined by \cite{Chu1965}, so that the streamwise oscillation of the disturbance is mainly absorbed in the phase part; ${\rm{Im}}(\cdot)$ is the imaginary part. Consequently, the second order, viscous derivative terms of the amplitude function in $\xi$ can be neglected. Introduction of the ansatz (\ref{ansatz}) in (\ref{LNS}) yields the plane-marching parabolized stability equations (PSE3D)
\begin{equation}\label{PSE}
\mathcal{P}\boldsymbol{\hat q} \equiv \left(\mathcal{A} + \mathcal{B}\frac{\partial}{\partial \xi}\right)\boldsymbol{\hat q} = \boldsymbol{\hat f},
\end{equation}
where $\mathcal{P}$ denotes the parabolized linearized Navier-Stokes operator (as detailed in \cite{Chen2022}) and $\boldsymbol{\hat f}$ is the Fourier component of the forcing at frequency $\omega$.
It is well known that some streamwise ellipticity remains in the streamwise pressure gradient term $\partial\hat p/\partial\xi$ and may cause the numerical instability. To eliminate the residual ellipticity, the streamwise pressure gradient term is either dropped from the equations as justified in \cite{Tempelmann2010} and \cite{Paredes2019} for stationary disturbances or replaced by $\Omega_{PNS}\partial\hat p/\partial\xi$, where $\Omega_{PNS}$ is the Vigneron parameter \citep{Vigneron1978}, for non-stationary disturbances.
\par
As criticized by \cite{Towne2019}, equation (\ref{alpha}) constrains the disturbances to have a unique, local wavenumber at any streamwise location such that PSE are likely to encounter difficulties when the perturbation field includes multiple modes with disparate streamwise wavenumbers or wavelengths. \cite{Towne2019} have shown that the errors introduced by the regularization techniques of PSE are proportional to the differences of the wavenmubers (or equivalently the distance between modes in the spectrum). Nevertheless, our local transient analysis results in section \S\ref{sec:3} indicate that the optimal growth perturbations primarily comprise modes concentrating in a small portion of the entire spectrum as is also observed by \cite{Hanifi1996}. This is reasonable since disturbances with close wavenumbers or phase velocities could efficiently interfere to achieve the optimal non-modal growth. Therefore, the absolute differences of the wavenumbers of these significant modes, and the errors pertaining to PSE shall be small. More detailed verifications can be found in Appendix \ref{Appendix}.

\subsection{Optimality system}
We seek the optimal response to the inlet disturbance or the external forcing. This problem can be readily formulated in an input-output framework below,
\begin{eqnarray}\label{I-O}
\mathcal{P}\boldsymbol{\hat q} = \mathcal{C}\boldsymbol{\hat f}, \\
\quad \boldsymbol{\hat\psi} = \mathcal{D}\boldsymbol{\hat q},
\end{eqnarray}
where $\boldsymbol {\hat f}$ denotes the input; $\boldsymbol {\hat \psi}$ designates the output; $\mathcal{C}$ and $\mathcal{D}$ are projectors (i.e., $\mathcal{C}\mathcal{C}=\mathcal{C}$, $\mathcal{D}\mathcal{D}=\mathcal{D}$) specifying how the input enters into the state equation and how the output extracts from the state $\boldsymbol{\hat q}$.
Define, respectively, the inner product and the energy-weight inner product
\begin{equation}
(\boldsymbol {\hat q},\boldsymbol {\hat q}) \equiv \int \boldsymbol {\hat q}^\dagger \boldsymbol {\hat q}{\rm{d}}\Omega, \quad (\boldsymbol {\hat q},\boldsymbol {\hat q})_E \equiv \int \boldsymbol {\hat q}^\dagger{\boldsymbol M}\boldsymbol {\hat q}{\rm{d}}\Omega, \quad (\boldsymbol {\hat f},\boldsymbol {\hat f})_F \equiv \int \boldsymbol {\hat f}^\dagger{\boldsymbol Q}\boldsymbol {\hat f}{\rm{d}}\Omega,
\end{equation}
where
\begin{equation}
\boldsymbol M \equiv {\rm{diag}}(\frac{\bar{T}}{\bar{\rho}\gamma Ma^2},\bar\rho,\bar\rho,\bar\rho,\frac{\bar\rho}{\gamma(\gamma-1)Ma^2\bar T})
\end{equation}
 represents the response energy weight, $\boldsymbol Q$ the forcing energy weight, $\Omega$ the domain where the shape function resides, $\dagger$ the transpose conjugate quantities. Two specific cases of $\mathcal{C}$ and $\mathcal{D}$ are considered in this paper. For the first case, we restrict the input as an impulse forcing at the inlet and the output as the state variables at the outlet such that the input and output behave like a Dirac function
 \begin{eqnarray}\label{Dirac}
 \mathcal{C}\boldsymbol{\hat f} = 0 \quad \text{except at the inlet}, \quad \int_{x_0}^{x_1}\mathcal{C}\boldsymbol{\hat f}{\mathbf{d}}x = \boldsymbol{\hat f}(x_0),\\
 \mathcal{D}\boldsymbol{\hat q} = 0 \quad \text{except at the outlet}, \quad  \int_{x_0}^{x_1}\mathcal{D}\boldsymbol{\hat q}{\mathbf{d}}x = \boldsymbol{\hat q}(x_1)
 \end{eqnarray}
 This case corresponds to the optimal response to the inlet perturbation. In the second case, we prescribe $\mathcal{C} = \mathcal{D} = \mathcal{I}$ (the identity operator) to retrieve the classical input-output problem where the forcing and response are distributed in the entire domain.
 \par
 The input and output can be related by the transfer (resolvent) operator $\mathcal{H}$ as
 \begin{equation}
 \boldsymbol{\hat\psi} = \mathcal{D}\mathcal{P}^{-1}\mathcal{C}\boldsymbol{\hat f} \equiv\mathcal{H}\boldsymbol{\hat f}.
 \end{equation}
 For the sake of simplicity, we assume $\boldsymbol {\hat q}$ to be a discretized vector such that we can neglect the integral symbol and assume that the forcing energy weight is the same as the response energy weight. As a result, the energy gain between the output and input reads
 \begin{equation}
 G=\frac{(\boldsymbol {\hat \psi},\boldsymbol {\hat \psi})_E}{(\boldsymbol {\hat f},\boldsymbol {\hat f})_F} = \frac{\boldsymbol {\hat f^\dagger}\mathcal{H}^\dagger\boldsymbol{M}\mathcal{H}\boldsymbol {\hat f}}{\boldsymbol {\hat f}^\dagger\boldsymbol M\boldsymbol {\hat f}} = \frac{\boldsymbol {\hat f}^\dagger\boldsymbol F^\dagger \boldsymbol F^{\dagger -1}\mathcal{H}^\dagger \boldsymbol F^\dagger \boldsymbol F\mathcal{H}\boldsymbol F^{-1}\boldsymbol F\boldsymbol {\hat f}}{\boldsymbol {\hat f}^\dagger\boldsymbol F^\dagger \boldsymbol F\boldsymbol {\hat f}},
 \end{equation}
 where $\boldsymbol M\equiv \boldsymbol F^\dagger \boldsymbol F$. The maximum energy gain reads
 \begin{equation}\label{twonorm}
 G_{max} = \left(\frac{\boldsymbol {\hat f}^\dagger\boldsymbol F^\dagger \boldsymbol F^{\dagger -1}\mathcal {H}^\dagger \boldsymbol F^\dagger \boldsymbol F\mathcal{H}\boldsymbol F^{-1}\boldsymbol F\boldsymbol {\hat f}}{\boldsymbol {\hat f}^\dagger\boldsymbol F^\dagger \boldsymbol F\boldsymbol {\hat f}}\right)_{max} = ||\boldsymbol F\mathcal{H}\boldsymbol F^{-1}||_2^2 = \sigma_1^2(\underbrace{\boldsymbol F\mathcal{H}\boldsymbol F^{-1}}_{\boldsymbol A}).
 \end{equation}
  $\sigma_1^2(\boldsymbol A)$ denotes the square of the maximum singular value of matrix $\boldsymbol A$, which equals the largest eigenvalue of the matrix $\boldsymbol A^\dagger \boldsymbol A$, and is readily solved by the power iteration with an arbitrarily given initial guess.
\subsubsection{Variational formulation}
 In practical computation, the above compact formulation (\ref{twonorm}) for optimal gain is usually adopted in local or two-dimensional temporal transient growth analysis when the matrix representation for $\mathcal{H}$ is available. For global non-modal analysis, the matrix-free method is preferred and it is customary to reformulate the optimal problem in a variational framework to facilitate calculation. The Lagrangian functional for this problem reads
 \begin{equation}\label{Lagrange2}
  J = (\mathcal{D}\boldsymbol{\hat q},\mathcal{D}\boldsymbol{\hat q})_E - {\rm{Re}}(\boldsymbol{\hat p},\mathcal{P}\boldsymbol{\hat q}-\mathcal{C}\boldsymbol{\hat f}) - \lambda\left((\mathcal{C}\boldsymbol{\hat f},\mathcal{C}\boldsymbol{\hat f})_F-1\right),
 \end{equation}
  where $\boldsymbol{\hat p}$, the Lagrange multiplier, is referred to as the adjoint variable; $\lambda$ is another Lagrange multiplier implementing the restriction on the input magnitude; ${\rm{Re}(\cdot)}$ is the real part.
  \par
  At the stationary point of the Lagrangian functional, we have, aside from the state equation (\ref{I-O}) and the input energy constraint $(\mathcal{C}\boldsymbol{\hat f},\mathcal{C}\boldsymbol{\hat f})_F = 1$, the adjoint equation
\begin{equation}\label{eig1}
\mathcal{P}^\dagger\boldsymbol{\hat p} = \mathcal{D}\boldsymbol M\mathcal{D}\boldsymbol{\hat q},
\end{equation}
and the optimal condition
\begin{equation}\label{optimal1}
\mathcal{C}\boldsymbol{\hat p} = \lambda \boldsymbol Q\mathcal{C}\boldsymbol{\hat f},
\end{equation}
which implies that the optimal forcing should be aligned with the adjoint state.
$\mathcal{P}^\dagger$ is the adjoint operator such that $(\boldsymbol{\hat p}, \mathcal{P}\boldsymbol{\hat q}) = (\mathcal{P}^\dagger\boldsymbol{\hat p},\boldsymbol{\hat q})$. Substitute (\ref{optimal1}) into (\ref{I-O}) to eliminate the forcing term $\boldsymbol{\hat f}$, we obtain
\begin{equation}\label{eig2}
\mathcal{P}\boldsymbol{\hat q} = \frac{1}{\lambda}\boldsymbol Q^{-1}\mathcal{C}\boldsymbol{\hat p}.
\end{equation}
Eqs. (\ref{eig1}) and (\ref{eig2}) form an eigenvalue problem which can be written as a single eigenvalue equation
\begin{equation}\label{RA-eig}
\mathcal{P^\dagger}^{-1}\mathcal{D}\boldsymbol{M}\mathcal{D}\mathcal{P}^{-1}\boldsymbol Q\mathcal{C}\boldsymbol{\hat p} = \lambda\boldsymbol{\hat p},
\end{equation}
where the leading eigenvalue $\lambda$ is exactly the optimal gain to the optimal forcing, and can be readily obtained by iteratively marching the direct and adjoint equations up to a converged state.
\par
For the optimal response to the inlet perturbation, we can derive an equivalent formulation that is more computationally tractable. Since $\mathcal{C}, \mathcal{D}$ satisfy (\ref{Dirac}), we can take integral with respect to $\xi$ in (\ref{I-O}) and let the outlet location $\xi_1$ approach the inlet location $\xi_0$, such that the forcing only enters into the inlet profile of the state variable, $\boldsymbol{\hat q_0}$, and is null downstream.
As such, Lagrangian functional (\ref{Lagrange2}) can be reformulated as
 \begin{equation}\label{Lagrange2}
  J = (\boldsymbol{\hat q_1},\boldsymbol{\hat q_1})_{ES} - {\rm{Re}}(\boldsymbol{\hat p},\mathcal{P}\boldsymbol{\hat q}) - \lambda\left((\boldsymbol{\hat q_0},\boldsymbol{\hat q_0})_{ES}-1\right),
 \end{equation}
where $\boldsymbol{\hat q_1}$ denotes the state profile at the outlet, $()_{ES}$ the energy-weight inner product over either the inlet or outlet. At the stationary point of $J$, we have the direct and adjoint equations
\begin{equation}\label{Adjoint}
 \mathcal{P}\boldsymbol {\hat q} = 0, \quad \mathcal{P}^\dagger\boldsymbol {\hat p} = 0,
\end{equation}
wherein $\mathcal{P}^\dagger \equiv \mathcal{A^\dagger} - \mathcal{B^\dagger}\frac{\partial}{\partial \xi}$ is the adjoint PSE3D (APSE3D) operator, and the optimal condition
  \begin{equation}\label{OptimalCondition}
\mathcal{B^\dagger}\boldsymbol {\hat p_1} = -\boldsymbol M\boldsymbol {\hat q_1}, \quad \mathcal{B^\dagger}\boldsymbol {\hat p_0} = \lambda \boldsymbol M\boldsymbol {\hat q_0}.
  \end{equation}
Again, the optimal growth can be readily obtained by iteratively solving the direct and adjoint equations with the optimal condition providing the inlet profiles (given an arbitrary initial guess). The optimal $N$-factor quantifying the integrated growth rate of the optimal disturbance is then defined as
 \begin{equation}
 N = \frac{1}{2}\ln(\lambda).
 \end{equation}\par


 \par
  The boundary conditions for the adjoint variables ($\boldsymbol {\hat p} = (\hat\rho_p,\hat u_p,\hat v_p,\hat w_p,\hat T_p)^{tr}$) are chosen so that the irrelevant boundary terms arising from the integration by parts vanish. At the wall, we have
  \begin{equation}
  \hat u = \hat v = \hat w = \hat T = 0, \quad \hat \rho_p = \hat u_p = \hat v_p = \hat w_p = \hat T_p = 0.
  \end{equation}
  At the upper boundary, we set
  \begin{equation}
  \hat u = \hat v = \hat w = \hat T = 0, \quad \hat u_p = \hat v_p = \hat w_p = \hat T_p = 0.
  \end{equation}
  Symmetry boundary conditions are implemented at the leeward and windward ray. Fourth-order finite difference scheme is used to discretize the azimuthal and wall-normal directions, while second-order Euler backward scheme is applied in the streamwise direction. The grid resolution of $200\times120$ for the azimuthal and wall-normal directions is used for the cases with 2 deg and 4 deg angles of attack, while the azimuthal points are increased to 300 for the largest AoA case. The solution was tested by changing the number of grid points and it was found to be grid-independent. Generally, four iterations suffice to get the converged results with the relative error regarding the optimal gain smaller than $10^{-3}$.
\subsubsection{Wavemaker}
One concept that we use in our subsequent analysis is that of structural sensitivity and the wavemaker \citep{Giannetti2007}. The wavemaker has its origin in global stability analysis and provides a way to highlight regions in which the global modes are most sensitive to the flow field changes (specifically, a localised flow feedback). Analogously, the resolvent wavemaker \citep{Skene2022} suggests the regions where flow field variations affectively change the optimal gain. Following the methodology of \cite{Schmid2014}, we consider base-flow perturbation of operators in  (\ref{eig1}) and (\ref{eig2}), which reads
\begin{eqnarray}\label{pert1}
\delta\mathcal{P^\dagger}\boldsymbol{\hat p} + \mathcal{P^\dagger}\delta\boldsymbol{\hat p} &=& \mathcal{D}\delta\boldsymbol M\mathcal{D}\boldsymbol{\hat q} + \mathcal{D}\boldsymbol M\mathcal{D}\delta\boldsymbol{\hat q}, \\ \label{pert2}
\delta\mathcal{P}\boldsymbol{\hat q} + \mathcal{P}\delta\boldsymbol{\hat q}  &=& -\frac{\delta\lambda}{\lambda^2}\boldsymbol Q^{-1}\mathcal{C}\boldsymbol{\hat p} + \frac{1}{\lambda}\boldsymbol Q^{-1}\mathcal{C}\delta\boldsymbol{\hat p}+ \frac{1}{\lambda}\delta(\boldsymbol Q^{-1})\mathcal{C}\boldsymbol{\hat p}.
\end{eqnarray}
Left multiplying (\ref{pert1}) by $\boldsymbol {\hat q^\dagger}$ and (\ref{pert2}) by $\boldsymbol {\hat p^\dagger}$, respectively, yields
\begin{eqnarray}\label{pert3}
(\boldsymbol {\hat q},\delta\mathcal{P^\dagger}\boldsymbol{\hat p}) + (\boldsymbol {\hat q},\mathcal{P^\dagger}\delta\boldsymbol{\hat p}) &=& (\boldsymbol {\hat q},\mathcal{D}\delta\boldsymbol M\mathcal{D}\boldsymbol{\hat q}) + (\boldsymbol {\hat q},\mathcal{D}\boldsymbol M\mathcal{D}\delta\boldsymbol{\hat q}), \\ \label{pert4}
(\boldsymbol {\hat p},\delta\mathcal{P}\boldsymbol{\hat q}) + (\boldsymbol {\hat p},\mathcal{P}\delta\boldsymbol{\hat q})  &=& -(\boldsymbol {\hat p},\frac{\delta\lambda}{\lambda^2}\boldsymbol Q^{-1}\mathcal{C}\boldsymbol{\hat p}) + (\boldsymbol {\hat p},\frac{1}{\lambda}\boldsymbol Q^{-1}\mathcal{C}\delta\boldsymbol{\hat p}) + (\boldsymbol {\hat p},\frac{1}{\lambda}\delta\boldsymbol Q^{-1}\mathcal{C}\boldsymbol{\hat p}).
\end{eqnarray}
Summing (\ref{pert3}) and (\ref{pert4}) to eliminate the terms associated with the perturbed shape functions, i.e., $\delta\boldsymbol{\hat q}$ and $\delta\boldsymbol{\hat p}$, by employing the direct and adjoint equations (\ref{eig1}) and (\ref{eig2}), leads to
\begin{equation}\label{pert5}
(\boldsymbol {\hat q},\delta\mathcal{P^\dagger}\boldsymbol{\hat p}) + (\boldsymbol {\hat p},\delta\mathcal{P}\boldsymbol{\hat q}) = (\boldsymbol {\hat q},\mathcal{D}\delta\boldsymbol M\mathcal{D}\boldsymbol{\hat q}) + (\boldsymbol {\hat p},\frac{1}{\lambda}\delta\boldsymbol Q^{-1}\mathcal{C}\boldsymbol{\hat p}) -(\boldsymbol {\hat p},\frac{\delta\lambda}{\lambda^2}\boldsymbol Q^{-1}\mathcal{C}\boldsymbol{\hat p}).
\end{equation}
By substituting (\ref{optimal1}) into (\ref{pert5}), and remembering the projector feature and the forcing amplitude constraint, (\ref{pert5}) can be further simplified to
\begin{equation}
2{\rm{Re}}(\boldsymbol {\hat p}, \delta\mathcal{P}\boldsymbol{\hat q}) = -\delta\lambda + (\boldsymbol{\hat q},\delta \mathcal{D}\boldsymbol M\mathcal{D}\boldsymbol {\hat q}) + (\boldsymbol {\hat p},\frac{1}{\lambda}\delta\boldsymbol Q^{-1}\mathcal{C}\boldsymbol{\hat p}),
\end{equation}
or equivalently
\begin{equation}
2{\rm{Re}}(\boldsymbol {\hat f}, \delta\mathcal{P}\boldsymbol{\hat q}) = \frac{-\delta\lambda}{\lambda} + \frac{1}{\lambda}(\boldsymbol{\hat q},\delta \mathcal{D}\boldsymbol M\mathcal{D}\boldsymbol {\hat q}) + (\boldsymbol {\hat p},\frac{1}{\lambda^2}\delta\boldsymbol Q^{-1}\mathcal{C}\boldsymbol{\hat p}).
\end{equation}

Therefore, we obtain the changes of the optimal gain to perturbation of base flow as
 \begin{eqnarray}
 \delta\lambda &=& -2 {\rm{Re}}(\boldsymbol{\hat p}, \delta\mathcal{P}\boldsymbol{\hat q}) + (\boldsymbol{\hat q},\delta \mathcal{D}\boldsymbol M\mathcal{D}\boldsymbol{\hat q}) + (\boldsymbol {\hat p},\frac{1}{\lambda}\delta\boldsymbol Q^{-1}\mathcal{C}\boldsymbol{\hat p})\\&=& -2 {\rm{Re}}(\boldsymbol{\hat f}, \delta\mathcal{P}\boldsymbol{\hat q})\lambda + (\boldsymbol{\hat q},\mathcal{D}\delta \boldsymbol M\mathcal{D}\boldsymbol{\hat q})+ (\boldsymbol {\hat p},\frac{1}{\lambda}\delta\boldsymbol Q^{-1}\mathcal{C}\boldsymbol{\hat p}).
 \end{eqnarray}
 The terms $(\boldsymbol{\hat q},\mathcal{D}\delta \boldsymbol M\mathcal{D}\boldsymbol{\hat q})$ and $(\boldsymbol {\hat p},\frac{1}{\lambda}\delta\boldsymbol Q^{-1}\mathcal{C}\boldsymbol{\hat p})$ simply reflect the changes related to the energy weight, while the first term of the right hand side indicates that the important structural-sensitivity region corresponds to the overlap regions of the adjoint variable (forcing) and the response. By neglecting the trivial terms pertaining to the energy weight, we can derive a bound for the gain drift under a unit-norm perturbation $\| \delta\mathcal{P}\| = 1$ as
 \begin{equation}
 |\delta\lambda| \leq 2\int \| \boldsymbol{\hat q} \|\| \boldsymbol{\hat p} \| {\rm{d}}\Omega = 2\lambda \int \| \boldsymbol{\hat q} \|\| \boldsymbol{\hat f} \|{\rm{d}}\Omega,
 \end{equation}
 where $\|\cdot\|$ should be understood as the pointwise norm of the mode. The wavemaker $\Lambda$ is then defined as
 \begin{equation}\label{wavemaker}
 \Lambda \equiv  \| \boldsymbol{\hat q} \|\| \boldsymbol{\hat p} \| = \lambda\| \boldsymbol{\hat q} \|\| \boldsymbol{\hat f} \|.
 \end{equation}
 In this way, the wavemaker $\mathbf{\Lambda}(\xi,\eta,\zeta)$ can be thought of as a field quantifying what changes to the optimal gain occur from localised feedback at each location. It can also be interpreted as a mechanism for self-sustained oscillations in the flow, since the optimal forcing (adjoint state) shows regions receptive to external perturbations and the optimal response reveals the structures being excited due to the forcing.
\par


\section{Results for the case of 2 degrees angle of attack}\label{sec:3}
\begin{figure}
\centering
\includegraphics[width = 0.9\textwidth]{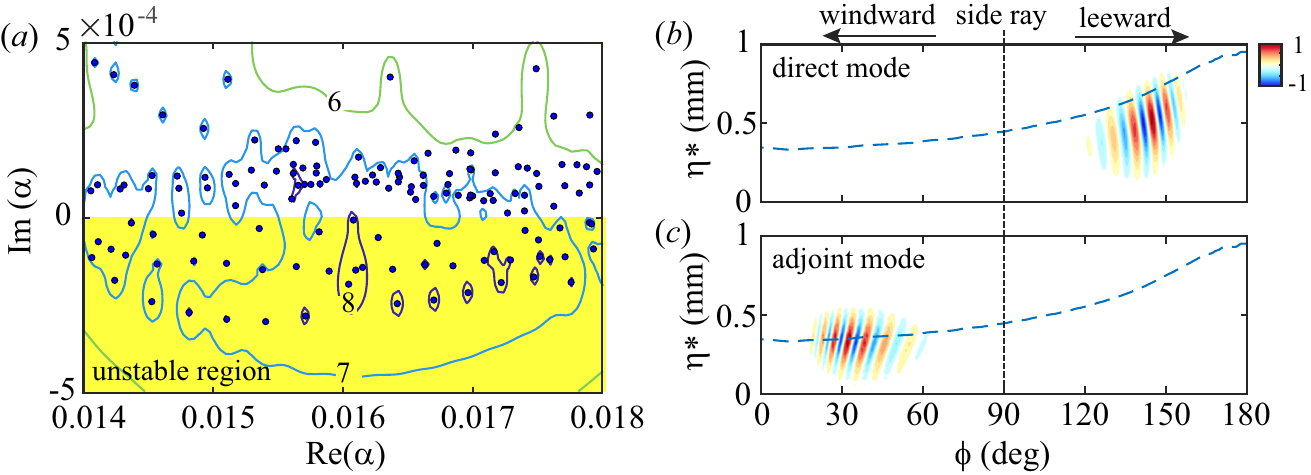}
\caption{($a$) Eigenspectrum (dots) and pseudospectrum (solid lines) of the linearized Navier-Stokes operator for frequency of 20 kHz at $X^* = 100$ mm. The isolines represent pseudospectrum given by $\log_{10}\epsilon^{-1}$ contours, with $\epsilon$ ranging from $10^{-8}$ to $10^{-6}$. The real parts of the normalized $\hat u$ shape function for the mode $\alpha = 0.0172-0.0002i$ and its adjoint are displayed in ($b$) and ($c$), respectively. The boundary-layer edge (dashed line) defined as the location for which the local total enthalpy equals 99\% of the freestream total enthalpy is also displayed.} 
\label{spectra}
\end{figure}
In this section, we will present results on the case of 2 degrees AoA, scrutinizing the salient features of the non-modal growth over a truly 3-D boundary layer, whereupon the results on the other AoA cases would be briefly discussed in the next section. As mentioned above, the crossflow modes are highly sensitive, which can be best illustrated by the pseudospectrum $\sigma_\epsilon(\boldsymbol A)$ \citep{Trefethen2005}. The pseudospectrum is defined by the set of eigenvalues ($z\in {\mathbb{C}}$) of a perturbed matrix ($\boldsymbol A$) such that
\begin{equation}
z\in\sigma(\boldsymbol A+\boldsymbol{\tilde e}),
\end{equation}
for some $\boldsymbol{\tilde e}$ with $\|\boldsymbol{\tilde e}\|<\epsilon$. The pseudospectrum can also be characterized alternatively by the resolvent norm as
\begin{equation}
\|(z-\boldsymbol A)^{-1}\|>\epsilon^{-1},
\end{equation}
which quantifies physically the maximum response of the flow to any forcing. When $\boldsymbol A$ is non-normal, the resolvent norm may retain extremely large values even far from the spectrum of $\boldsymbol A$. A representative pseudospectrum visualized by contours of the resolvent norm in the logarithmic scale for frequency of 20 kHz at $X^* = 100$ mm is shown in figure \ref{spectra}($a$). Many unstable modes can be observed, as expected, and they almost all lie within the contour line of $ \|(z-\boldsymbol A)^{-1}\| = 10^{7}$, meaning that one can not exactly distinguish any of these unstable modes under perturbations as small as $10^{-7}$. Such a high sensitivity of the eigenvalues obviously stems from the non-normality of the operator which can in turn cause a substantial transient growth.
\par

The non-normality of the operator can be further inferred by inspecting figure \ref{spectra}($b,c$), showing the spatial distribution of a typical direct-adjoint mode pair. The direct mode manifests as a tilted wavepacket structure lying on the leeward region, whereas the adjoint mode leaning in the opposite direction resides on the windward side. Such a clear separation in space between direct and adjoint modes indicates high sensitivities of the eigenvalues and a strong azimuthal non-normality induced by crossflow \citep{Chomaz2005}.
\par
\subsection{Optimal response to inlet perturbation}
\begin{figure}
\centering
\includegraphics[width = 0.9\textwidth]{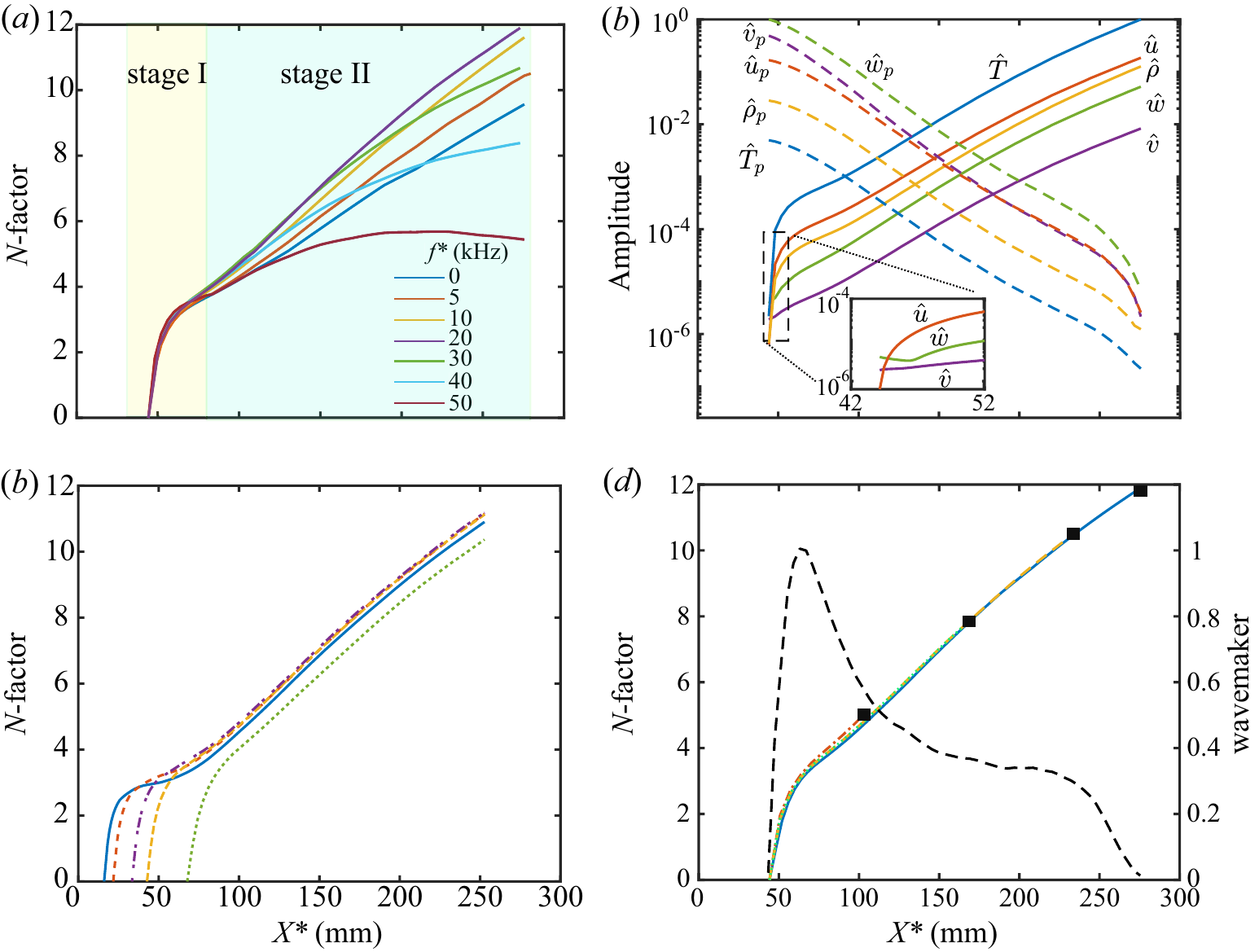}
\caption{Parametric effects on the $N$-factors of the optimal growth: ($a$) different frequencies with $X_0^* = 44$ mm and $X_1^* = 275$ mm; ($b$) streamwise evolution of the component amplitude of the direct (solid lines) and adjoint (dashed lines) states for frequency of 20 kHz; ($c$) different inlet locations with a fixed outlet location ($X^*_1 = 250$ mm) and a frequency of 20 kHz; ($d$) different outlet locations (marked by the square symbol) with a fixed inlet location ($X^*_0 = 44$ mm) and a frequency of 20 kHz, along with the normalized wavemaker distribution ($\max_{\eta,\zeta}\Lambda(\xi,\eta,\zeta)$, dashed line).} 
\label{AoA2-N-f-x}
\end{figure}
As a first step we perform a parametric study on the optimal growth regarding the inlet/outlet positions and the disturbance frequency. The results are displayed in figure \ref{AoA2-N-f-x} from which we can make several observations. First, the optimal disturbances undergo two distinct evolution stages: they initially experience a short-term, yet substantial transient growth (achieving growth of $\delta N\approx4$ within about 20 mm)(stage I) followed by a relatively long-term moderate growth (stage II). The disturbance frequency seemingly affects the second stage only. The most amplified frequency is around 20 kHz whose amplitude evolution of the direct and adjoint variables are shown in figure \ref{AoA2-N-f-x}($b$). It can be seen that the temperature fluctuation amplitude rapidly grows and soon becomes dominant, followed by the streamwise velocity component ($\hat u$). In contrast, the adjoint variables exhibit an opposite trend, suggesting a strong receptivity of the response to streamwise vortices ($\hat v, \hat w$) in the vicinity of the inlet. Second, there exists an optimal initial position given a frequency and an outlet position where the seeded disturbance can amplify maximally. For frequency of 20 kHz, the optimal inlet location is at around $X^* = 40$ mm. At last, extending the outlet position with a given initial position does not modify the path but only raises the final energy gain reached by the disturbance. This can be explained by the wavemaker position. By definition (\ref{wavemaker}), the wavemaker should lie in the overlap region of the direct and adjoint states. Since the adjoint variable amplitude decays at a smaller rate in the vicinity of the inlet than the growth rate of the direct mode near the outlet, the wavemaker concentrates in a small region downstream the inlet. Therefore, the optimal disturbance characteristics would be substantially preserved as long as the instability core (wavemaker) is included inside the spatial interval.
\par
\begin{figure}
\centering
\includegraphics[width = 0.9\textwidth]{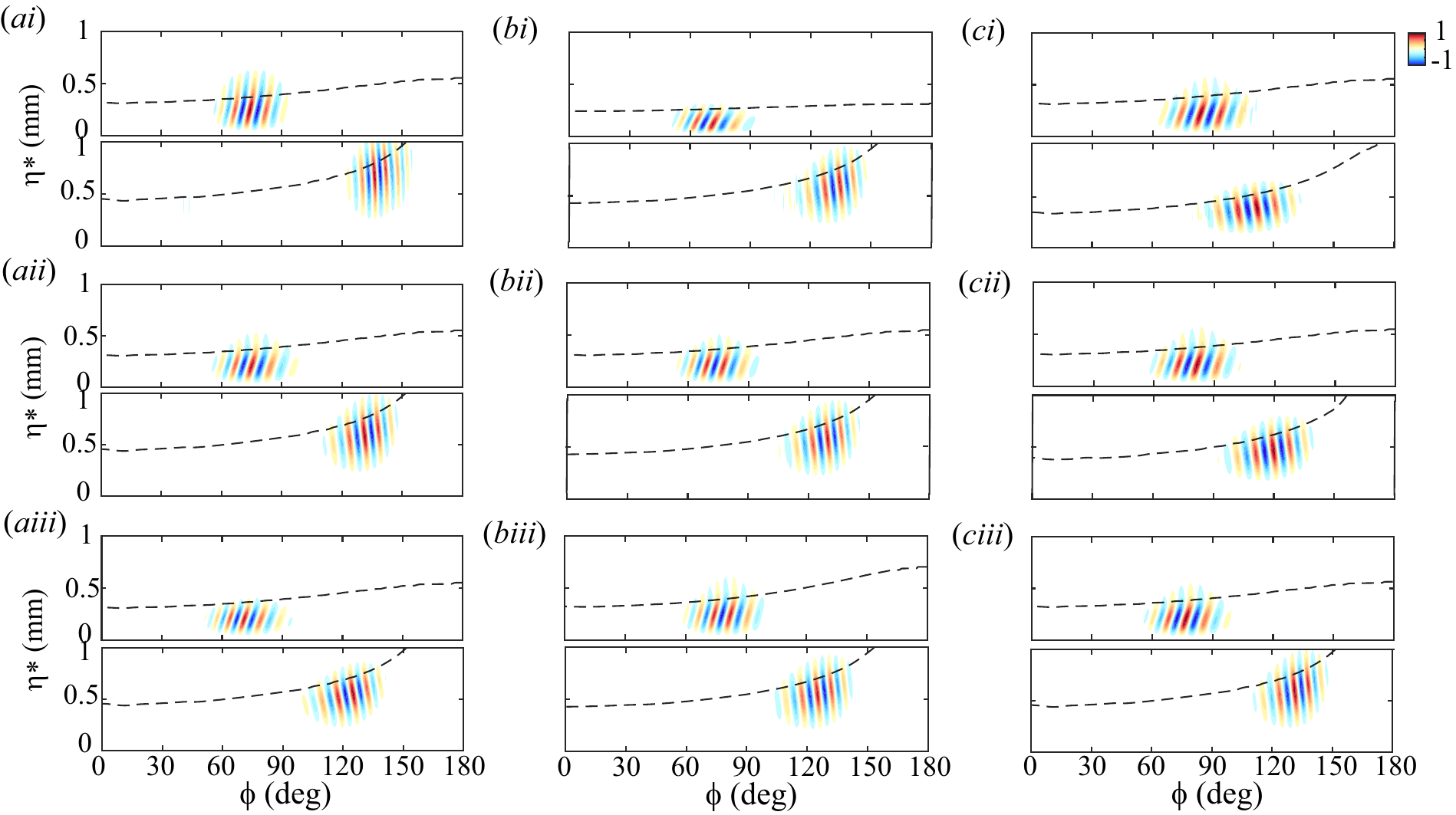}
\caption{Parametric effects on the optimal disturbance profiles depicted by the real part of the temperature component at the inlet and outlet positions (shown in the upper and lower row, respectively, within each subplot). Left column: varying frequency for a fixed spatial domain (same as in figure \ref{AoA2-N-f-x}($a$)), ($ai$) stationary disturbance, ($aii$) 20 kHz, ($aiii$) 40 kHz. Middle column: varying inlet positions for a given frequency (20 kHz) and a fixed outlet position ($X_1^* = 250$ mm), ($bi$) $X_0^*  = 16$ mm, ($bii$) $X_0^* = 43$ mm, ($biii$) $X_0^* = 68$ mm. Right column: varying outlet positions for a given frequency (20 kHz) and a fixed inlet position ($X_0^* = 44$ mm), ($ci$) $X_1^* = 103$ mm, ($cii$) $X_1^* = 234$ mm, ($ciii$) $X_1^* = 275$ mm. The dashed line denotes the boundary-layer edge. } 
\label{AoA2-profile-f-x}
\end{figure}
The parametric effects on the optimal disturbance profiles at the inlet and outlet locations are illustrated in figure \ref{AoA2-profile-f-x}. One can observe that the optimal disturbances all initially manifest as a tilted wavepacket lying preferentially on the windward side, very reminiscent of the shape function of a typical adjoint mode as displayed in figure \ref{spectra}. At the outlet position, they all tilt in the opposite direction and reside on the leeward side like a crossflow mode. With increasing disturbance frequency, the wavepacket is initially more tilted and ultimately resides closer to the side ray ($\phi = 90$ deg). On the other hand, the initial wavepacket progressively approaches the windward ray for a more upstream inlet, yet appears to be insensitive to the outlet location. This observation is important because one could obtain the optimal initial profile for even the entire model through performing optimal calculation over a relatively small domain where the wavemaker is strong.
\par
\begin{figure}
\centering
\includegraphics[width = 0.8\textwidth]{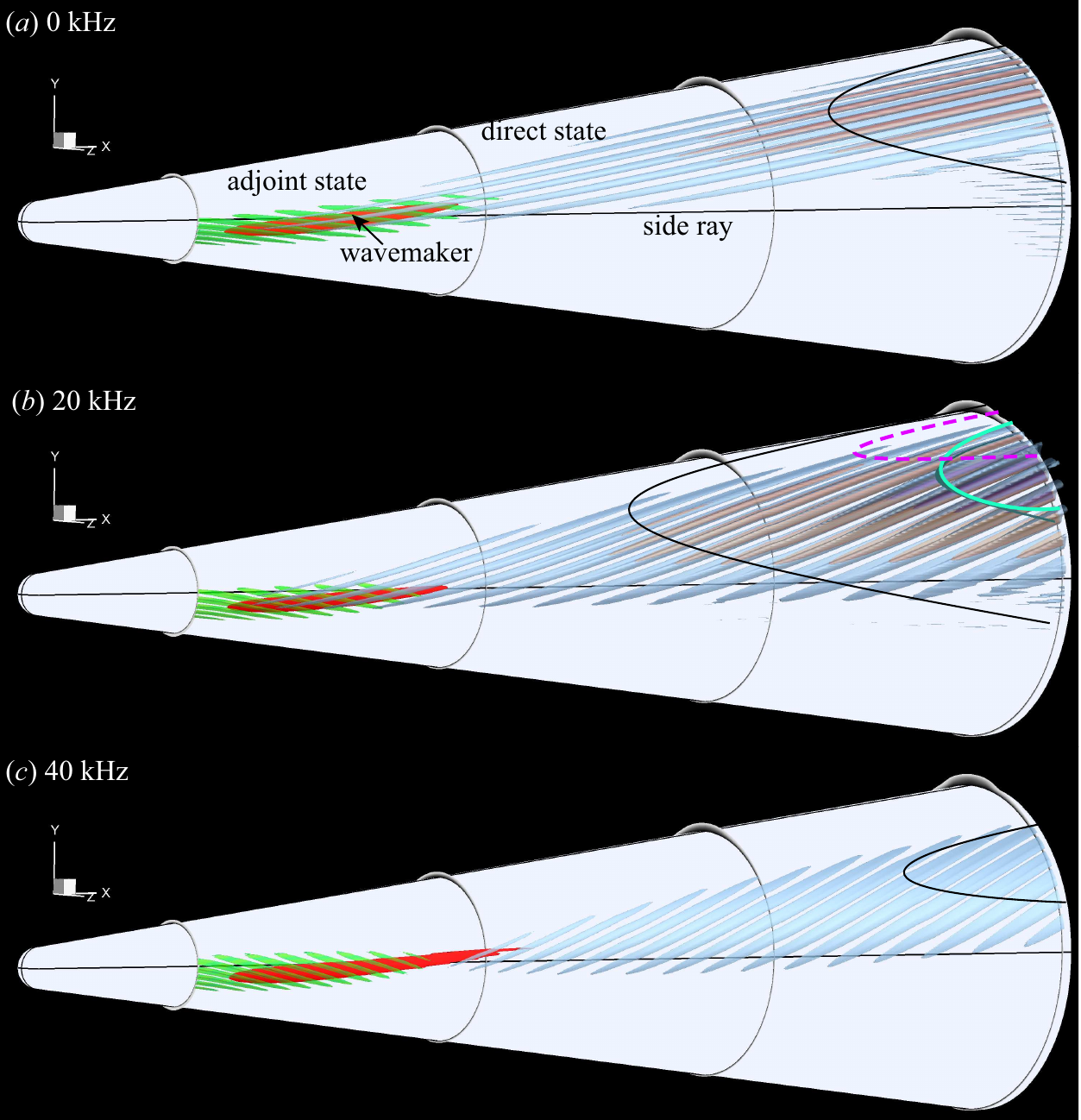}
\caption{Spatial structures of the optimal disturbances of three representative frequencies, visualized by the isosurfaces of the real part of the temperature shape function with amplitude envelopes corresponding to $N_T = 5$ (blue), $N_T = 8$ (red) and $N_T = 11$ (purple). The temperature shape function is normalized by the maximum value at the inlet position. ($a$) $f^* = 0$ kHz, ($b$) $f^* = 20$ kHz, ($c$) $f^* = 40$ kHz. Streamwise velocity distribution at four successive streamwise slices ($\eta$-$\zeta$ plane) starting from $X^* = 44$ mm to $X^* = 275$ mm is also shown. The solid lines denote the $N$-factor envelope calculated from the LST-eN method (black: $N_T = 5$, cyan: $N_T = 8$). The dashed pink line in ($b$) denotes the disturbance amplitude envelope corresponding to $N_T = 2.3$ obtained from PSE3D initiated by modal disturbances at 20 kHz. The spatial structures of the adjoint state (green, {\rm{Re}}($\hat T_p$) = 0.1$|\hat T_p|_{max}$) and the wavemaker (red, 0.5$|\Lambda|_{max}$) are also displayed.} 
\label{AoA2-tecplot}
\end{figure}
Figure \ref{AoA2-tecplot} shows the spatial structures of direct and adjoint states for three frequencies. We observe that the direct disturbance starts on the windward side and moves towards the leeward side, exhibiting a triangle-like amplitude envelope embodying several bent streaks inside; with increasing frequency, the core part of the spatial structure moves progressively towards the windward side, and the streaks are more prominently tilted. On the other hand, the adjoint fluctuation, quantifying the high-receptivity region, lies preferentially on the windward side close to the inlet, exhibiting wave angles with respect to the streamwise direction opposite to those of the direct disturbances. Such remarkable spatial separation of the direct and adjoint states is indicative of the streamwise as well as azimuthal non-normality \citep{Chomaz2005}. Interestingly, the wavemaker of whatever frequency, marking the structural sensitivity region, manifests as a longitudinal structure slightly downstream from the inlet on the windward side. Consequently, one may expect to employ some control strategy in the wavemaker region for the crossflow instabilities with any frequency.

\par
The area occupied by the amplitude envelope is obviously proportional to the $N$-factor of the disturbance. In analogous to the classical eN method utilizing the 1-D stability analysis (LST), we define the local $N$-factor distribution on the cone surface based on the temperature disturbance amplitude as
\begin{equation}
N_T(\xi,\zeta) = \ln(\max_\eta |\hat T(\xi,\eta,\zeta)|/\max_{\eta,\zeta}|\hat T(\xi_0,\eta,\zeta)|).
\end{equation}
Note that a similar formulation was adopted by \cite{Araya2023} with wall pressure as the alternative measure of disturbance amplitude. As such, the amplitude envelope of the optimal disturbance provides an upper bound of the transition front pertaining to a given $N$-factor in the eN framework. Figure \ref{AoA2-tecplot} compares the transition fronts corresponding to $N_T = 5$, 8 and 11 (as long as the targeted $N_T$ can be reached) predicted by the non-modal analysis and by modal analyses based on LST (LST-eN) and on PSE3D tracing every unstable global mode of 20 kHz at $X^* = 100$ mm downstream (PSE3D-eN). The first $N$-factor ($N_T = 5$) is typically associated with laminar-turbulent transition of stationary crossflow vortices in the incompressible regime \citep{Bippes1999}, while the second one ($N_T = 8$) is close to the transition $N$-factor of traveling crossflow waves on the Mach 6 HyTRV model \citep{Chen2022}. Evidently, the optimal disturbance amplitude envelopes significantly extend in the upstream direction the LST-eN envelope at the same $N$-factor as well as the PSE3D-eN envelope even with a much smaller $N$-factor ($N = 2.3$), highlighting the appreciable non-modal growth. Nevertheless, modal analyses also predict the predominance of the 20 kHz waves, suggesting that non-modal growth does not change the peak frequency. Another interesting observation is that the core part of the downstream optimal structure matches well with the most unstable region predicted by the LST-eN method, but is remarkably far away from the leeward ray compared to the global modal amplitude envelope. This implies that the optimal disturbance does exploit the crossflow instability, yet not in the same manner as a global crossflow mode does, which will be discussed in detail later.
\par

\subsubsection{Growth mechanism}
\begin{figure}
\centering
\includegraphics[width = 0.9\textwidth]{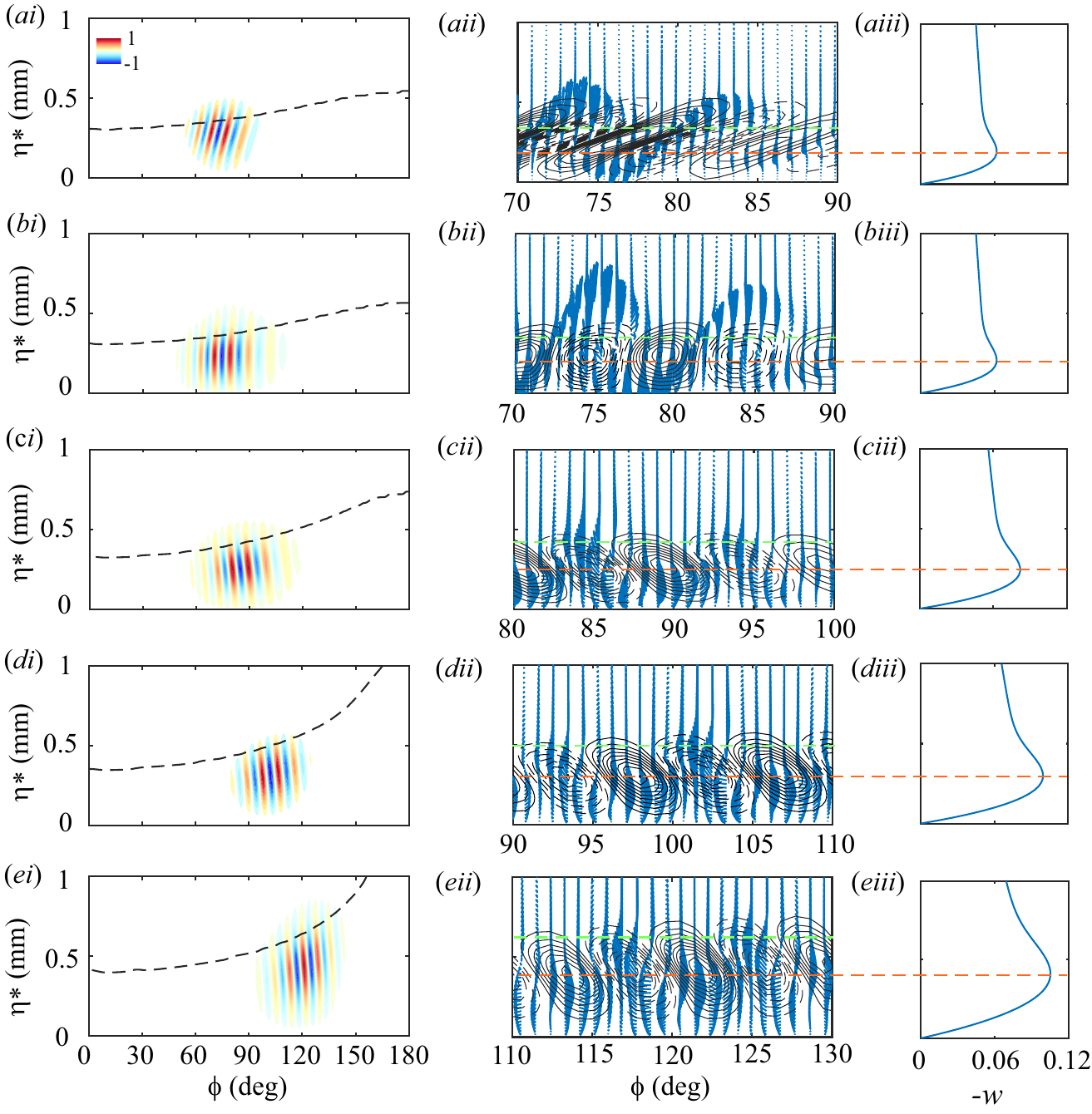}
\caption{Visualization of the downstream evolution of the optimal disturbance at frequency of 20 kHz optimized in the spatial domain of 44 mm $\leq X^* \leq$ 275 mm. Shown on the left column is the streamwise velocity contours at five successive stations: ($a$) $X^* = 44$ mm denoting the initial position, ($b$) $X^* = 48$ mm, ($c$) $X^* = 72$ mm, ($d$) $X^* = 119$ mm and ($e$) $X^* = 200$ mm. The dashed line denotes the boundary-layer edge. Shown on the middle column is the zoomed-in of the main part of the optimal disturbance in the left column with the vector plot comprising the wall-normal and crossflow velocity fluctuations. The mean crossflow profile at the peak location of the optimal disturbance is shown in the right column.} 
\label{AoA2-initialprofile}
\end{figure}
The non-modal growth mechanisms can be inferred from downstream evolution of the velocity profiles as shown in figure \ref{AoA2-initialprofile}. First, we observe an azimuthal shift of the disturbance structure from the windward side to the leeward side as the optimal disturbance travels downstream. Such a spatial movement originates in the azimuthal non-normality \citep{Chomaz2005}, and is well known for the temporal optimal disturbance in 2-D boundary layers, who superposed by streamwise global modes is convected from upstream to downstream \citep{Ehrenstein2005,Alizard2007,Akervik2008}. As opposed to the 2-D cases where the optimal disturbance gains energy through the T-S instability mechanism during the upstream-downstream propagation process, the crossflow instability mechanism is mainly responsible for the disturbance energy growth in the present azimuthal movement.
\par
Moreover, the tilted direction of the disturbance structures changes rapidly at the first three stations, suggesting the Orr mechanism \citep{Butler1992}. By analogous to the analysis of \cite{Butler1992}, the Orr mechanism in the present case is rooted in the production term, $-{\rm{Re}}(\hat w^\dagger \hat v)\frac{\partial \bar w}{\partial \eta}$, stemming from the azimuthal momentum equation. When this term is positive, or equivalently the disturbance structure is titled against the shear of the crossflow, the disturbance is able to gain energy from the mean flow. Otherwise, the disturbance would lose energy. Figure \ref{AoA2-initialprofile} reveals that the crossflow profile exhibits two distinct regions across the boundary layer with opposite sign pertaining to the shear term ($\frac{\partial \bar w}{\partial \eta}$). As a result, if the disturbance gains energy in the upper region, it would return energy back to the base flow in the lower region with the same tilted direction. At the initial position, the disturbance structure well extends into the free stream and is tilted against the upper shear, favoring the energy gain. The slanted structure soon turns upright at the next station and tilts along the opposite direction further downstream. Beyond the second station, the disturbances penetrate more deeply into the boundary layer, occupying almost equally the two shear regions of the crossflow, and the total energy transfer pertaining to the Orr mechanism should thus be smaller thereafter.
\par
\begin{figure}
\centering
\includegraphics[width = 0.7\textwidth]{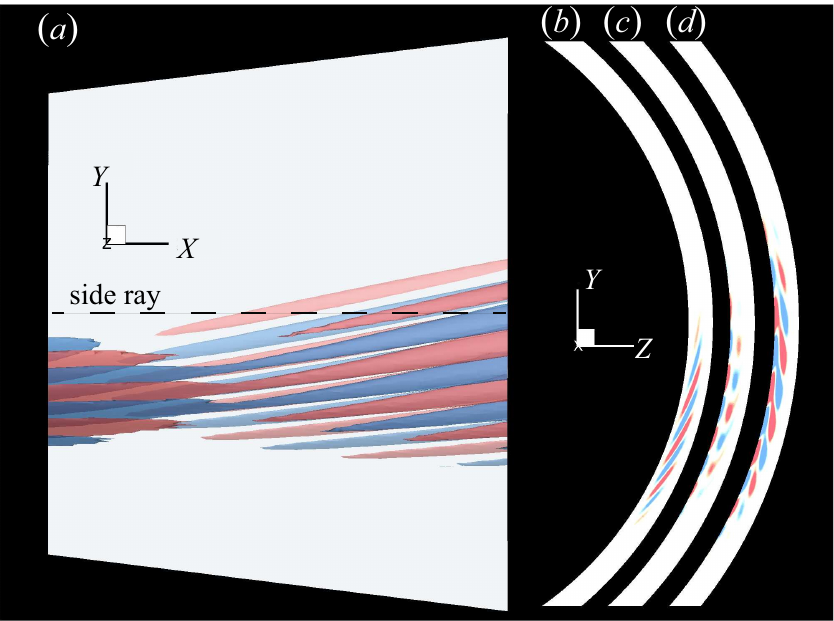}
\caption{Optimal structure evolution in stage I (44 mm $<X^*<64$ mm) depicted by ($a$) the isosurfaces of the axial vorticities (red: positive, blue: negative) and the axial vorticity distribution on the cross sections at three successive axial stations ($b$) $X^* = 44$ mm, ($c$) $X^* = 54$ mm, ($d$) $X^* = 64$ mm. The cross section coordinates ($Y,Z$) have been enlarged by twice for facilitating comparison.} 
\label{AoA2-f20-initial}
\end{figure}
Figure \ref{AoA2-f20-initial} clearly delights how the optimal profile initiates a crossflow-instability wavepacket, from which an important insight on the physical mechanisms in stage I can be gained. It can be seen that the optimal profile gives rise to immediately downstream alternating positive and negative longitudinal vortical rolls with axes aligned along the axial direction, which are well known in shear-layer flows to be able to optimally leverage the lift-up effects. The lift-up effect can also be inferred from the observation that the initially dominant vortex components ($\hat v,\hat w$) are soon taken over by the streak component ($\hat u$) in amplitude, as displayed in figure \ref{AoA2-N-f-x}($b$).  Despite weakening rapidly due to viscous dissipation, these longitudal vortices induce counter-rotating vortices below, forming two-cell vorticity structures with axes being tilted with respect to the axial direction. Such vorticity structures are no longer optimal regarding the lift-up effect because of the counteracting forces from two opposite-orientation vorticity layers, yet are typical of crossflow instabilities \citep{Wassermann2003} of which they can take advantage to sustain the downstream growth. At the same time, the apparent change of the vorticity inclination in the cross section as shown in figure \ref{AoA2-f20-initial}($b$-$d$) is indicative of the Orr mechanism that can enhance the lift-up effect as well as the near-wall vorticity induction by strengthening the primary vorticity.
This vortex evolution process is representative of optimal disturbances, the quantitative difference being that the inclined angle of the two-cell vorticity structures increases with the frequency. As per the observations made so far we might thus conclude that the non-modal growth of the optimal disturbance in stage I primarily stems from a synergistic combination of the lift-up effect and the Orr mechanism, which conforms the optimal growth results for the infinite swept wing boundary layer (SWBL) \citep{Tempelmann2010,Tempelmann2012}.
\par

\begin{figure}
\centering
\includegraphics[width = 0.9\textwidth]{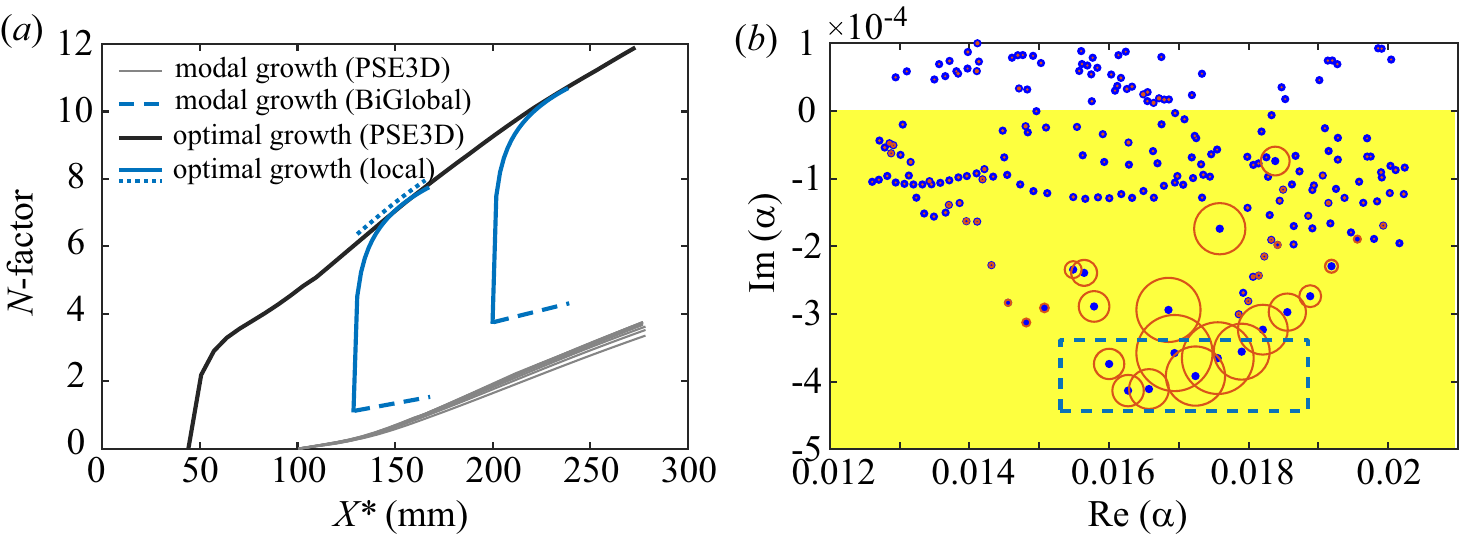}
\caption{($a$) $N$-factor comparison of the optimal disturbance at frequency of 20 kHz and the crossflow modal disturbances obtained from PSE3D initiated by different crossflow modes at $X^* = 100$ mm at the same frequency (grey lines). The results from the local transient growth analysis (solid lines) and the most unstable mode obtained by BiGlobal (dashed lines) for two representative stations ($X^* = 134$ mm and 200 mm) are displayed, with an upward shift to facilitate comparison. ($b$) Contributions of individual global modes to the local optimal transient growth at $X^* = 134$ mm, visualized by the circles sized proportional to the corresponding expansion coefficients $\kappa_k$. The energy gain of the disturbance superimposed by few most unstable modes (delimited by the dashed box) is also shown in ($a$) (dotted line). } 
\label{AoA2-modalandnonmodal}
\end{figure}

In the following, we consider whether the crossflow-instability wavepacket initiated by the optimal profile is governed by a dominating unstable mode as in the SWBL \citep{Tempelmann2010}. Figure \ref{AoA2-modalandnonmodal}($a$) compares the $N$-factor of the optimal disturbance of 20 kHz and those calculated by PSE3D with initial profiles provided by different crossflow modes (see figure \ref{spectra}($a$) for spectrum) at $X^* = 100$ mm (deemed close to the neutral location). Obviously, the optimal disturbance at stage II amplifies at a substantially larger rate than the modal disturbances do, indicating that it could not be described properly by a single mode. This is not surprise because the global crossflow instability spectrum as shown in figure \ref{AoA2-modalandnonmodal}($b$) contains multiple unstable modes without a predominant one, in stark contrast to the SWBL case where generally only one unstable crossflow mode exists given a frequency and a wavenumber. The (linear) interference among these similarly-looking modes may yield a much more rapid energy growth than the most unstable mode, which could have been exploited by the optimal disturbance. To justify such speculation, we have conducted the local transient growth analysis, seeking an optimal combination of modes that reaches the maximum energy growth at the specific distance downstream, for two representative stations. In the framework of the local transient analysis, the disturbance is expanded into $n$ eigenmodes
\begin{equation}\label{transient}
\boldsymbol q'(t,\xi,\eta,\zeta) = \exp(-i\omega t)\sum_{k=1}^n\kappa_k\boldsymbol {\hat q_k} (\eta,\zeta)\exp(i\alpha_k\xi).
\end{equation}
The vector-function $\boldsymbol {\hat q_k}$ corresponds to the $k$th eigenfunction. The coefficients $\kappa_k$ are optimized to achieve the maximum energy growth at the specific downstream coordinate. The maximum energy gain can be cast as the same form as (\ref{twonorm}). Thanks to the ansatz (\ref{transient}), the resolvent or transfer matrix $\mathcal{H}$ can be put explicitly as follows \citep{Tumin2001}:
\begin{equation}
\mathcal{H} = {\rm{diag}}(\exp(i\alpha_1\xi),\exp(i\alpha_2\xi),\dots,\exp(i\alpha_n\xi)),
\end{equation}
and accordingly, the entries of the weighting matrix $\boldsymbol M_{k,l} = (\boldsymbol {\hat q_k},\boldsymbol {\hat q_l})_E$. The resulting maximum energy gain (\ref{twonorm}) can be determined by using the singular value decomposition.
\par
The spectrum used in calculating the local optimal energy gain for $X^* = 134$ mm is displayed in figure \ref{AoA2-modalandnonmodal}($b$). It contains a total of 208 modes, but only a small portion of the spectrum primarily comprising significantly unstable modes plays an important role. The corresponding optimal energy gain, shown in figure \ref{AoA2-modalandnonmodal}($a$) with an artificial upward shift to facilitate the comparison, clearly reveals the initial algebraic-growth stage and the ensuing quasi exponential-growth stage. It should be noted that the local transient growth analysis fails to capture initial energy gain due to the Orr and lift-up mechanisms which might need much more modes than presently used \citep{Bagheri2009}. Nevertheless, the growth rate pertaining to the quasi exponential-growth stage is substantially larger than that of the most unstable mode, and bears a striking resemblance to that of the global optimal disturbance for an appreciably long distance. The optimal energy gain based on few most unstable modes (delimited by the dashed box) exhibits a diminishing algebraic growth, yet quantitatively the same exponential growth as the fuller spectrum. This indicates that the quasi exponential-growth stage is governed by the few most unstable modes while the less unstable modes would affect the initial algebraic-growth stage only. Such observations seem to hold at other stations as well, justifying the speculation that the optimal disturbance may have exploited the interference among few most unstable modes locally to achieve the global optimal growth. Indeed, the global transient analysis on 2-D flow configurations has shown that superposition of non-normal global (even decay) modes recovers the convective (exponential) instability \citep{Cossu1997,Bagheri2009,Nichols2011}. The difference is that the disturbances are allowed to amplify only in the streamwise direction in the 2-D flows, whereas they can gain energy along both the streamwise and azimuthal directions in the present 3-D configuration.
\par
\begin{figure}
\centering
\includegraphics[width = 0.9\textwidth]{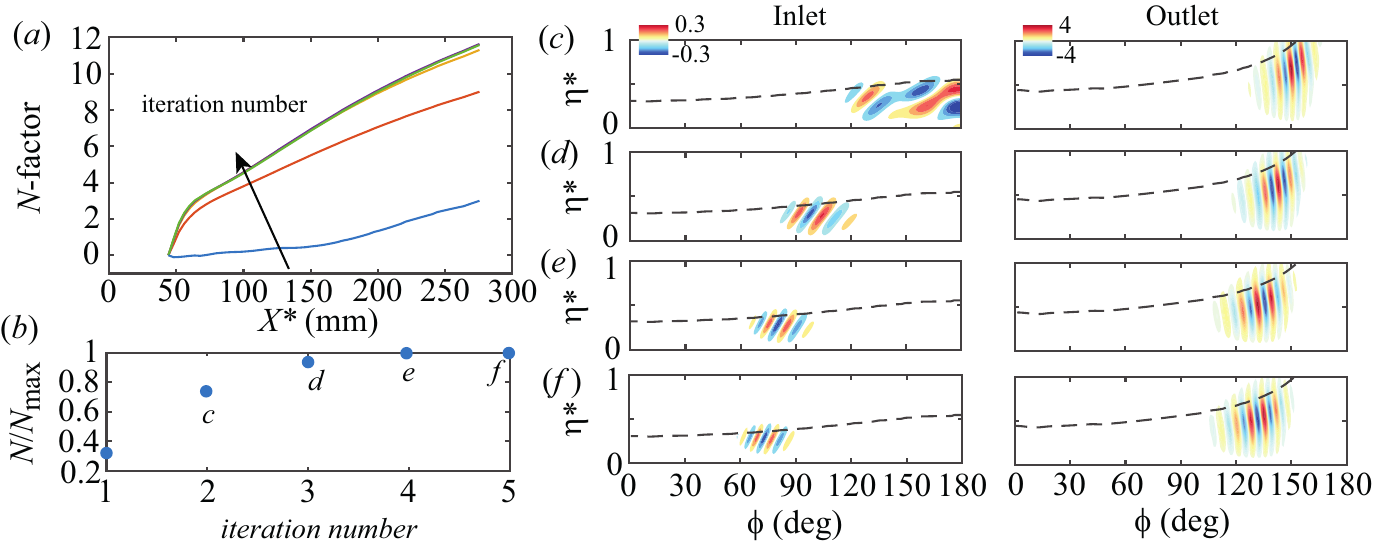}
\caption{Convergence process of the optimal disturbance at 20 kHz in the streamwise range of $(44,275)$ mm initiated by a random profile: ($a$) the $N$-factor evolution, ($b$) the ratio of the outlet $N$-factor to the optimal one; ($c$-$f$) streamwise velocity profiles normalized by the cross velocity components ($\sqrt{\hat v^2 + \hat w^2}$) at the inlet (middle column) and outlet (right column) for the 2nd iteration through 5th iteration, with the dashed line denoting the boundary-layer edge.} 
\label{AoA2converg}
\end{figure}
The above results reveal three active mechanisms beneath the optimal growth, i.e., the Orr mechanism, the lift-up effect, and the crossflow instability mechanism manifesting both in the streamwise and azimuthal directions. A significant insight regarding the role played by these mechanisms can be gained from the typical convergence process of the optimal growth, as displayed in figure \ref{AoA2converg}. We observe that a regular disturbance concentrated in the leeward ray already emerges from the random initial profile after just one forward-backward iteration, retrieving all the aforementioned mechanisms and yielding nearly 80 percent of the final optimal energy gain (figure \ref{AoA2converg}($b$)): its streawise velocity component is much smaller than the vortex component in amplitude, an indicator of the lift-up effect; the slanted structures in the $\eta$-$\zeta$ plane suggest the Orr mechanism; at last, the outlet profile exhibiting typical crossflow-instability features implies the crossflow-instability mechanism. After one more iteration, the inlet disturbance structure entirely leans against the crossflow shear such that it can take full advantage of the Orr mechanism; moreover, the spatial separation is more pronounced, enhancing the disturbance energy gain along the azimuthal direction through crossflow instability. Further iteration drives the inlet disturbance to the windward side progressively, yet barely improves the optimal gain, indicating that the azimuthal crossflow-instability growth is relatively weak compared to other mechanisms. In summary, the iteration process tends to first filter out the lift-up effect and the streamwise crossflow instability mechanism, which can thus be viewed as most important for the optimal growth, and then fully exploits the Orr and azimuthal crossflow instability mechanisms to achieve the maximum amplification.
\par
\subsubsection{High-frequency optimal growth}
\begin{figure}
\centering
\includegraphics[width = 0.9\textwidth]{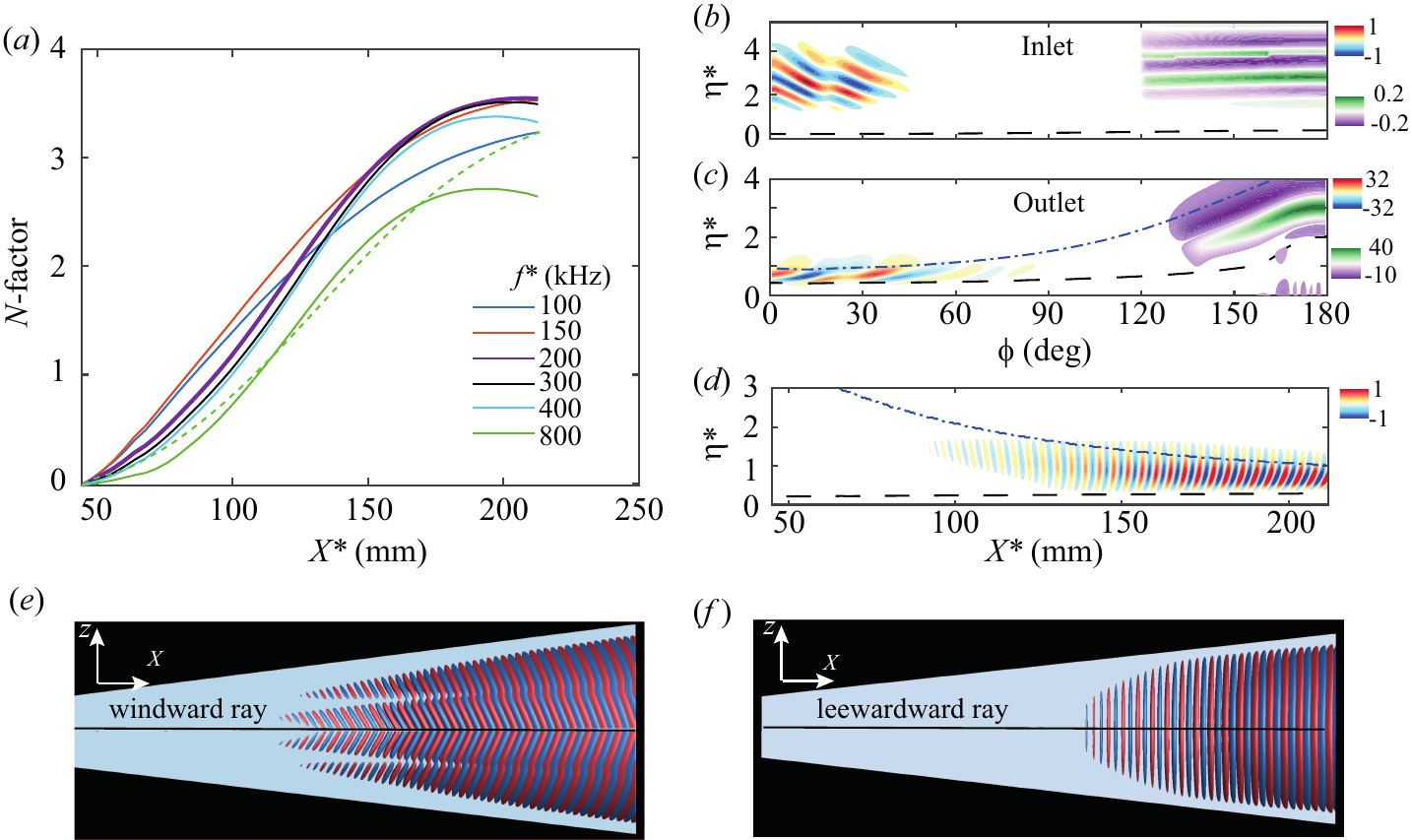}
\caption{Results for optimal disturbances of high frequencies optimalized over a streamwise interval corresponding to $x\in(44,212)$ mm. ($a$) Evolution of $N$-factor of various frequencies. The dashed line denotes the result for the 200 kHz optimalized for the restricted azimuthal region $\phi\in (120, 180)$ deg. ($b$-$d$) Contours of the normalized temperature of the optimal disturbance at 200 kHz displayed in the inlet ($b$) and outlet ($c$) as well as in the $X$-$\eta$ plane at the windward ray ($\phi$ = 0 deg). The temperature components at ($b$) and ($c$) are normalized by the total velocity amplitude, while those at ($d$) are normalized by the maximum temperature amplitude. The blue-red contours and the green-purple contours denote the optimal structure optimalized for the entire azimuthal region and for the restricted azimuthal region, respectively. The dashed and dashed-dotted lines indicate the edge of the boundary layer, $\delta_h$, and the edge of the entropy layer, $\delta_S$, respectively. ($e$) Spatial structure evolution for the full optimal disturbance at 200 kHz depicted by the isosurfaces of the temperature disturbance (red: positive, blue:negative, amplitude corresponding to $N_T$=3.2).} 
\label{AoA2highf}
\end{figure}

In addition to the low-frequency optimal disturbances, we also calculated optimal disturbances at high frequencies which exhibit totally different features from the former. Figure \ref{AoA2highf} depicts the evolution of $N$-factors of frequencies ranging from 100 kHz to 800 kHz for a specified streamwise interval and the temperature contour of one representative frequency. One can observe moderate transient growths, albeit much weaker than the low-frequency optimal growth, experienced by a broad-band disturbance far from the modal instability frequencies (around 1 MHz for Mack mode instabilities). The relative differences of the peak $N$-factors among various frequencies are less than 1, indicating a relatively flat frequency amplification. The optimal shape function, exemplified by the frequency of 200 kHz, concentrates in the domain between the boundary-layer edge and the entropy-layer edge of the windward side throughout the entire streamwise interval. Following the work of \cite{Paredes2020}, the entropy-layer edge, $\delta_S$, is defined as the location where the local entropy increment is 0.25 times the entropy increment at the wall ($\Delta \bar S(\eta=\Delta_S) = 0.25\Delta \bar S_{wall}$). The entropy increment, $\Delta \bar S$, is defined with respect to the freestream value, i.e., $\Delta \bar S = 1004 \ln (\bar T)-287\ln (\bar P)$. The fact that the disturbance remains in the vicinity of the windward ray indicates that the aforementioned azimuthal shift of the low-frequency optimal disturbances is solely due to the crossflow instability, which is absent here. The mode shapes at the inlet and outlet positions exhibit opposite slanted structures in both the azimuthal and streamwise directions, suggesting an Orr-like mechanism responsible for the transient growth. As opposed to the classical Orr mechanism which mainly amplifies the velocity components, the temperature component of the high-frequency perturbation is the prime benefactor (see figure \ref{AoA2highf}($b,c$)) from algebraic amplification where the temperature gradient within the entropy layer is a driving factor. Specifically, the optimal disturbance is capable of extracting energy from the mean temperature shear in the wall-normal and azimuthal directions through terms $-{\rm{Re}}(\hat T^\dagger \hat v)\frac{\partial \bar T}{\partial \eta}$ and $-{\rm{Re}}(\hat T^\dagger \hat w)\frac{\partial \bar T}{\partial \zeta}$, respectively. Finally, the spatial evolution of the optimal disturbance at 200 kHz is illustrated in figure \ref{AoA2highf}($e$), featuring a wave train structure in the vicinity of the windward ray with a preferential wave angle on either side. Such wave structure is reminiscent of that of the Mack mode instability on a similar configuration \citep{Chen2023PRF}, and is thus presumably to trigger the boundary layer breakdown via the same mechanism, i.e., the (generalized) fundamental resonance.
\par
It is noteworthy that \cite{Paredes2019} have also discovered entropy-layer traveling disturbances that experience appreciable non-modal growth on a hypersonic blunt cone at zero AoA. They \citep{Paredes2020} later showed that a pair of oblique non-modal waves could induce onset of transition via oblique breakdown. Our work extends their results to a 3-D configuration, showing that the optimal entropy-layer disturbances concentrate in the vicinity of the windward ray. Because the optimal disturbance on either side of the windward ray comprises exclusively components with wave angles of the same sign, prohibiting interactions between components with opposite wave angles, the classical oblique breakdown mechanism is not active herein. It is interesting to assess whether high-frequency disturbances can undergo transient growth in other places than the windward ray as well. A straightforward method is to calculate the optimal disturbance within a restricted azimuthal domain. As inferred from figure \ref{AoA2highf}, the optimal disturbance restricted in the leeward region, exhibiting an essentially 2-D structure, turns out to enjoy a commensurate energy amplification as well through exploiting an Orr-like mechanism in the streamwise direction (not shown here).
\subsection{Optimal response to external forcing}
\begin{figure}
\centering
\includegraphics[width = 0.9\textwidth]{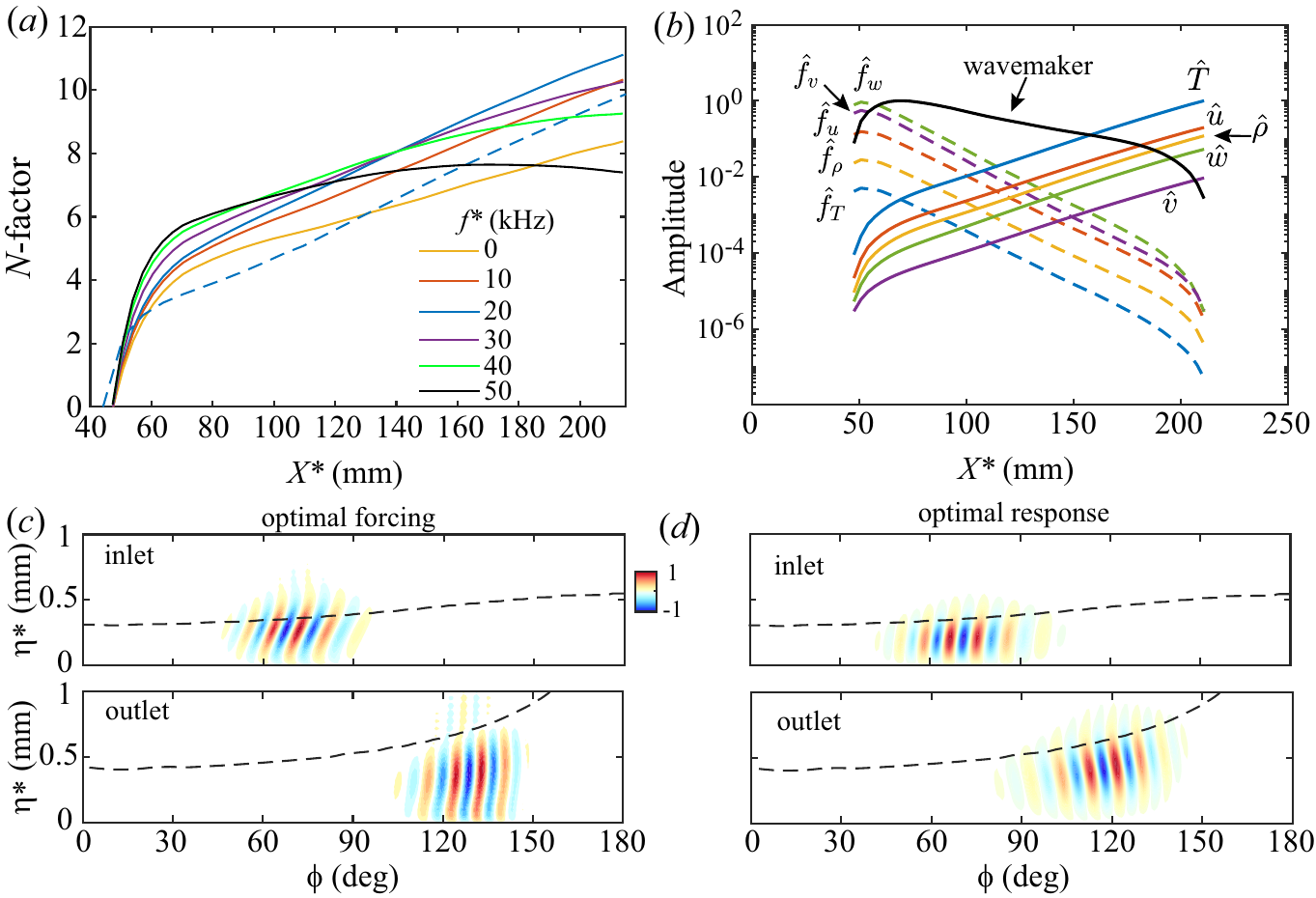}
\caption{($a$) $N$-factor of the optimal response to external forcings at frequency ranging from $f^* = 0$ to $f^* = 50$ kHz, together with the optimal response to inlet profiles at 20 kHz (dashed line) for comparison. ($b$) Streamwise component amplitude evolution of the normalized optimal response (solid lines) and the normalized optimal forcing (dashed lines) for 20 kHz. The normalized resolvent wavemaker distribution ($\max_{\eta,\zeta}\Lambda(\xi,\eta,\zeta)$) is also displayed. The corresponding optimal forcing and response shape functions at the inlet (upper row) and outlet (lower row) positions, illustrated by $\hat f_u$ and $\hat u$, are plotted in ($c$) and ($d$), respectively } 
\label{f20NR}
\end{figure}
In this subsection, we present results about optimal response to the external forcing in the entire domain. Without loss of generality, we set $\boldsymbol Q = \boldsymbol I$ in measuring the forcing energy. Because the inlet response is prescribed null for this case, the conventional $N$-factor would yield an infinitely large value. To circumvent such difficult, we consider the integrated growth rate between a position slightly downstream the inlet and the outlet. Figure \ref{f20NR} shows the results for a representative spatial interval $X^*\in(44,215)$ mm. The $N$-factors starting from $X^* = 47$ mm in figure \ref{f20NR}($a$) feature an algebraic growth stage followed by a long-term (quasi exponential-growth) phase, and peak at 20 kHz, closely resembling the optimal growth results. Direct comparison between the dominant $N$-factor evolution for the optimal forcing case and the optimal inlet-profile case indicates that their differences lie primarily in the algebraic-growth stage wherein the former is remarkably enhanced. Interestingly, the algebraic growth is more pronounced with increasing frequency, in stark contrast to the optimal inlet-profile case wherein it almost coincides for different frequencies.
\par
The streamwise distribution of the component amplitude of the optimal forcing and the response for 20 kHz is displayed in figure \ref{f20NR}($b$). As expected, the forcing amplitudes exhibit an almost opposite trend against the response counterpart; they peak within the algebraic-growth stage and rapidly diminish downstream. Forcing terms on the $\zeta$- and $\eta$-direction momentum equations are dominant, implying that the optimal forcing tends to leverage the lift-up effect to prompt the energy growth by enhancing the streamwise vorticty. Moreover, the optimal forcing shape function shown in figure \ref{f20NR}($c$) is always tilted against the crossflow shear, suggesting that the optimal forcing exploits the Orr mechanism as well. The resulting response shown in figure \ref{f20NR}($d$) is very reminiscent of that in the optimal inlet-profile case. To conclude, the optimal disturbance under the external forcing undergoes essentially the same evolution process with the same growth mechanisms as in the optimal inlet-profile case; the optimal forcing mainly acts on the algebraic-growth stage (stage I) by constantly enhancing the lift-up and Orr mechanisms, yet barely affects the ensuing stage where the crossflow instability prevails.
\par
\begin{figure}
\centering
\includegraphics[width = 0.8\textwidth]{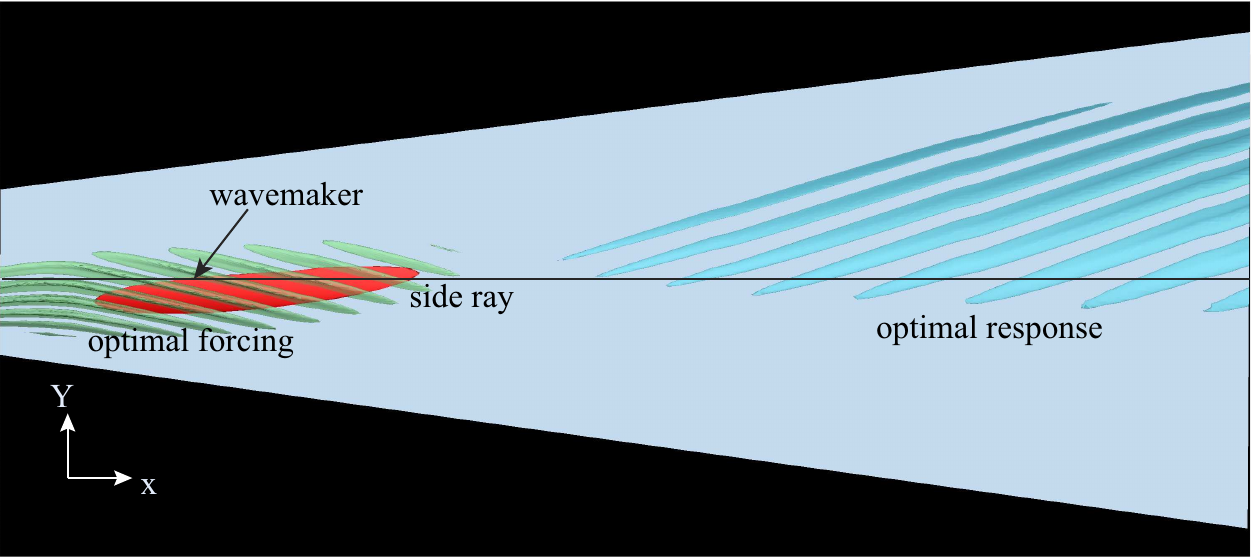}
\caption{Spatial structure of the optimal forcing (green, ${\rm{Re}}$($\hat f_T$)$=10^{-5}$), the optimal response (blue, ${\rm{Re}}$($\hat T) = 50000$) and the resolvent wavemaker (red, 0.5$|\Lambda|_{max}$).} 
\label{f20NR-structure}
\end{figure}


As inferred from figures \ref{f20NR-structure} and \ref{AoA2-tecplot}, the optimal forcing is exerted on the same position of the adjoint state except for the vicinity of the inlet where it manifests as longitudinal structures aligned with the axial direction. Since axially aligned vortices (with nearly zero streamwise wavenumber) are most favorable for the lift-up effect \citep{Farrell1993}, the optimal forcing presumably takes more advantage from the lift-up effect, thus enhancing the algebraic growth. Moreover, the wavemaker almost coincides with that of the optimal growth, implying the same localized feedback mechanisms for both cases.

\section{Influence of the angle of attack}\label{sec:4}
\begin{figure}
\centering
\includegraphics[width = 0.9\textwidth]{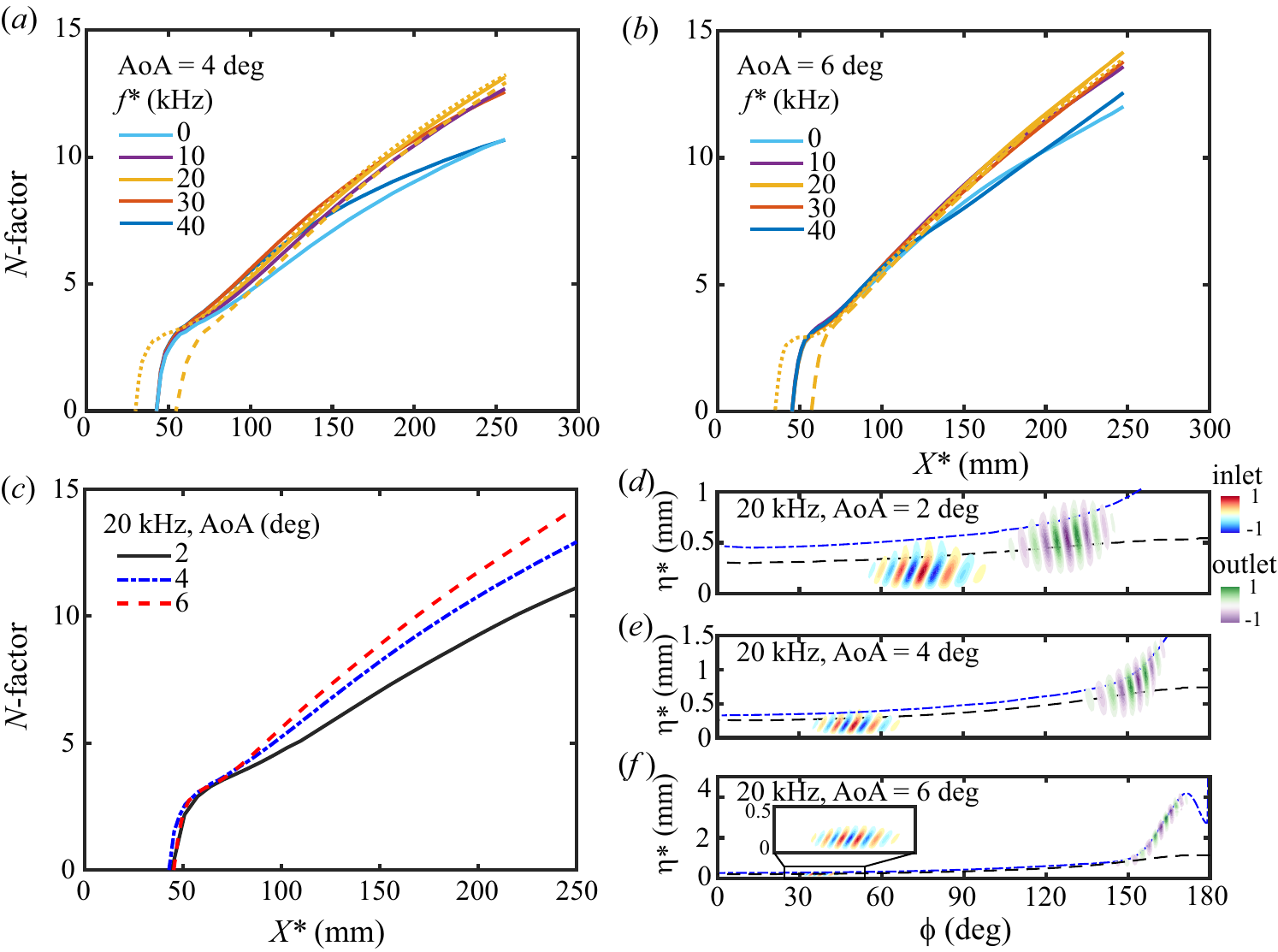}
\caption{$N$-factor comparison of the optimal disturbance at frequency ranging from $f^* = 0$ to $f^* = 40$ kHz for cases of 4 degrees AoA ($a$) and of 6 degrees AoA ($b$), and at frequency 20 kHz for all three cases ($c$). The dashed lines in ($a$) and ($b$) show the results obtained with suboptimal inlet positions. The corresponding real part of the temperature disturbance component at the inlet (red-blue contour with dashed lines denoting the boundary-layer edge) and outlet (green-purple contour with dashed-dotted lines denoting the boundary-layer edge) are displayed in ($d$-$f$), respectively. } 
\label{AoAeffect-N}
\end{figure}
In this section, we will assess the influence of the angle of attack on the non-modal growth. The $N$-factors of optimal disturbances to inlet perturbation for different frequencies in the case of 4 degrees AoA and the case of 6 degrees AoA are displayed in figure \ref{AoAeffect-N}($a$) and ($b$), respectively. For whatever AoA, the optimal inlet position remains at around $X^* = 44$ mm where the optimal disturbance achieves the largest energy growth, and the outlet position beyond the strong wavemaker region does not modify the evolution path except for raising the final energy gain reached by the disturbance (not shown here). Therefore, we only present the optimal growth results for the spatial interval of $X^*\in(44,250)$ mm here. Notably, the optimal disturbances in all three cases share qualitatively the same evolution path consisting of the algebraic-growth stage (stage I) insensitive to the disturbance frequency and the following quasi exponential-growth stage (stage II) with growth rates relying on the disturbance frequency. Moreover, the most amplified frequency is always around 20 kHz. As shown in figure \ref{AoAeffect-N}($c$), increasing the angle of attack does not change the growth rate in the algebraic-growth stage, but considerably enhances the disturbance amplification in stage II in accord with the strengthening of the crossflow velocity and thus the crossflow instability. Moreover, a larger AoA yields a more pronounced separation of the inlet and outlet disturbance profiles in the azimuthal direction (figure \ref{AoAeffect-N}($d$-$f$)), with the inlet disturbance closer to the windward ray and the outlet disturbance closer to the leeward ray. This obviously stems from the stronger crossflow velocity pertaining to a larger AoA and implies an increasingly important role played by the azimuthal crossflow instability mechanism in the optimal growth.
\par
\begin{figure}
\centering
\includegraphics[width = 1.0\textwidth]{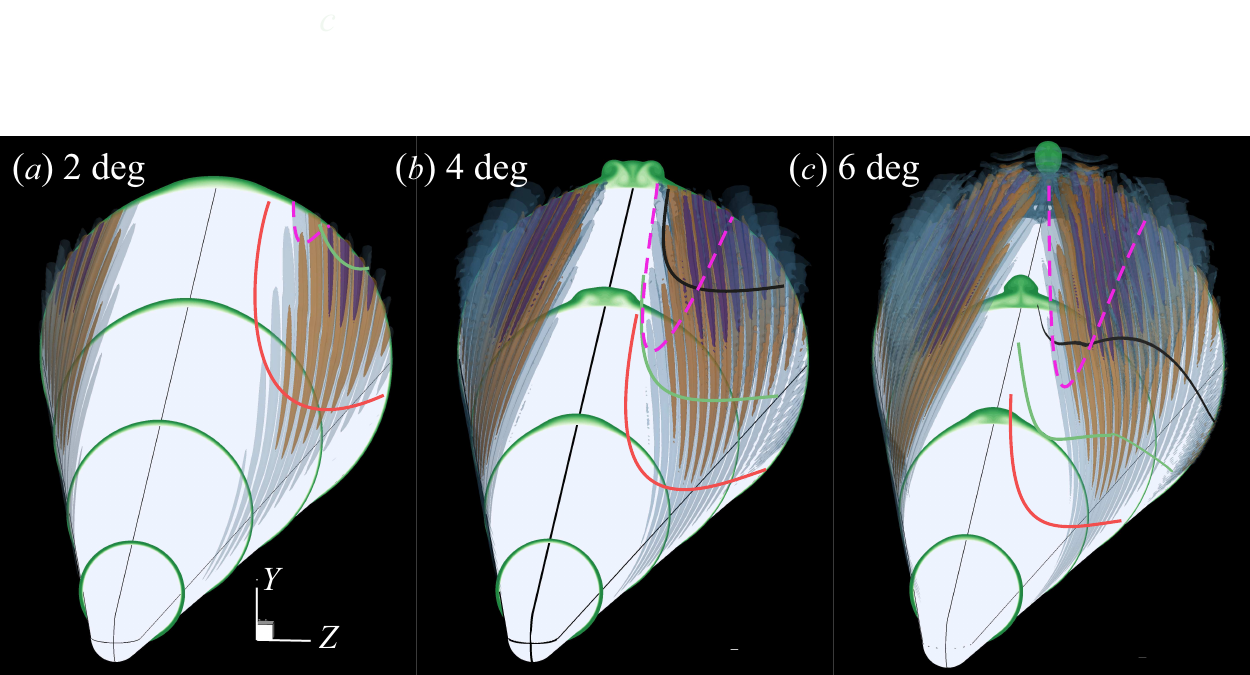}
\caption{Spatial structures of the optimal disturbance at 20 kHz for three angles of attack, visualized by three isosurfaces, $N_T = 5$ (blue), $N_T = 8$ (orange) and $N_T = 11$ (purple), of real part of temperature disturbances normalized by the maximum value at the initial profile for each case, together with the corresponding three $N$-factor envelopes ($N = 5$ red, $N = 8$ green, $N = 11$ black) obtained from the LST-eN method and that ($N=2.3$ pink dashed line) initiated by the most amplified global crossflow mode obtained from PSE3D. Contours of axial velocity at four successive axial locations starting from the inlet position are also displayed.} 
\label{AoAeffect2}
\end{figure}
The spatial structures of the most amplified optimal disturbances for three cases are compared in figure \ref{AoAeffect2}, featuring the tilted streaks on the cone surface with larger inclined angles for increasing AoA. Intriguingly, the optimal disturbance at 6 deg AoA has penetrated partly into the leeward vortex region, which likely interferes or even precipitates the transition process therein as documented by \cite{Wang2023} in a numerical study on a similar flow configuration. \cite{Quintanilha2022} reported that the temporal optimal crossflow disturbances on a hypersonic elliptic cone ultimately reside in the vortex region, manifesting as intrinsic vortex modes. In this study, however, we find that the optimal disturbance preserves the crossflow-instability feature towards the outlet rather than develops into an intrinsic vortex instability.
\par
It is instructive to compare the optimal amplitude envelopes with the modal counterpart calculated respectively by PSE3D and LST. All methods predict stronger disturbance energy growth with increasing AoA, as expected. Again, the optimal disturbance is remarkably more amplified than the global modal counterpart, but such superiority quickly diminishes as the AoA increases. This is because increasing AoA greatly promotes the modal instability yet barely affects the algebraic growth, and as such, the modal instability becomes increasingly important. The $N$-factors based on LST are significantly larger than those of global modes and are even commensurate with the optimal ones at high AoAs. Aside from the $N$-factor, the amplitude envelopes from the three also differ in terms of the peak position in the azimuthal direction: the global modal amplitude envelope lies closest to the leeward ray, followed by the local one and the global optimal one.
\par
\begin{figure}
\centering
\includegraphics[width = 0.95\textwidth]{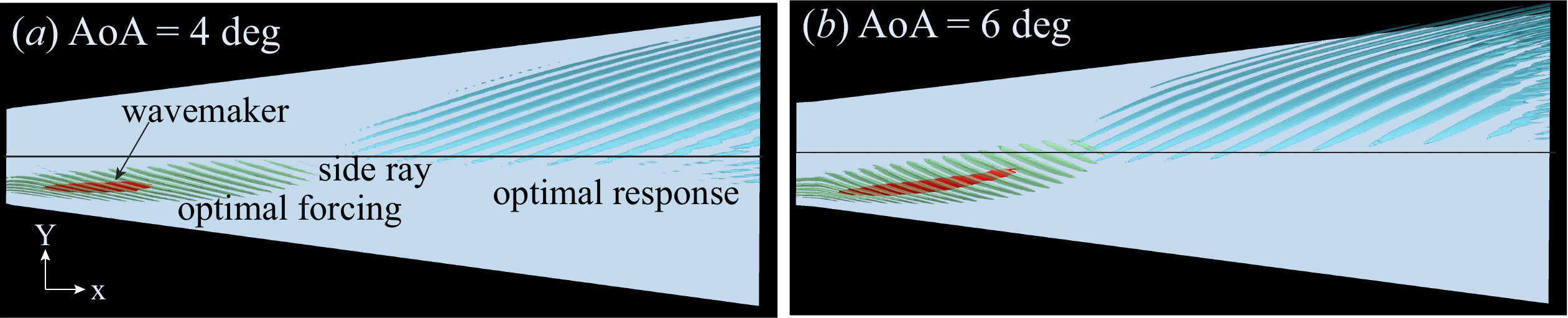}
\caption{Spatial structure of the optimal forcing (green, energy equation forcing real($\hat f_T$)$=10^{-5}$), the optimal response (blue, temperature component real($\hat T) = 50000$) and the resolvent wavemaker (red, 0.5$|\Lambda|_{max}$) for two angles of attack: ($a$) 4 degrees, ($b$) 6 degrees. } 
\label{AoA46NR}
\end{figure}
In summary, increasing the AoA simply enhances the modal instability, resulting in a more pronounced azimuthal shift of the disturbance, yet exerts relatively small influence on the initial algebraic growth. This observation also holds true for the optimal response to external forcing. For brevity, we only present the spatial structures of the optimal forcing/response and the wavemaker for the two higher AoA cases in figure \ref{AoA46NR}, showing essentially the same pattern as in the case of 2 deg AoA in figure \ref{f20NR-structure}.
\section{Conclusions and discussions}\label{sec:5}
In this work, the optimal responses to the inlet disturbances and the external forcing in the entire domain are investigated within a uniform input-output framework for hypersonic boundary layers on a blunt cone at nonzero angles of attack (AoAs). Well-established parabolized governing equations (PSE3D and APSE3D) are employed to efficiently obtain the spatial evolution of direct as well as adjoint states, which allows for a systematic parameter study. Comparisons with relevant literature data and other elaborate numerical tools including HLNS and DNS demonstrate the reliability of the present solver in calculating the non-modal disturbances. \par
The optimal growth analysis seeking the optimal response to inlet perturbation is first conducted for the case with 2 degrees AoA, which lays the foundation for finding the optimal response to external forcing and subsequent study on the effects of AoA. The optimal disturbances, peaking at 20 kHz, initially take the form of azimuthally compact tilted streamwise vortices on the windward side. While convected downstream, they quickly rise to an upright position and give rise to bent streaks, accompanied by a substantial energy growth. The initial phase of energy growth is algebraic, attributed to a combination of the lift-up effect and the Orr mechanism, and is later smoothly replaced by an exponential-like growth primarily driven by the crossflow instability mechanism. Such a two-stage growth process bears a strong resemblance to what has been observed by \cite{Tempelmann2010,Tempelmann2012} in an infinite swept-wing like boundary layer (SWBL). However, we note two important differences caused by the azimuthal inhomogeneity in the present configuration. Firstly, instead of following a single crossflow mode as in the SWBL case, the optimal disturbance in the second stage behaves more like a non-modal one comprised of several leading global crossflow modes, and thus amplifies much faster than the most unstable crossflow mode does. Secondly, the optimal disturbances are localised in the azimuthal direction rather than uniformly distributed in the SWBL case, and shift azimuthally from the windward side towards the leeward ray while traveling downstream such that they can exploit the crossflow instability mechanism not only along the streamwise direction but also along the azimuthal direction. This spatial movement of the disturbance structure prevails in the 2-D streamwise global non-modal analyses and is known to stem from the combination of multiple global modes \citep{Chomaz2005}. Increasing the AoA barely affects the algebraic-growth stage, yet substantially promotes the exponential growth through strengthening the crossflow instability, leading to a more prominent azimuthal shift of the disturbance structure.
 \par
 The structural sensitivity (wavemaker) of the optimal gain is shown to be quantified by the pointwise product between the direct and adjoint (or forcing) states. Our results indicate that the wavemakers of low-frequency optimal disturbances always reside on the windward side immediately downstream of the inlet position where efficient transition control might be implemented.
 \par
 The optimal external forcing concentrates in the vicinity of the inlet with amplitude rapidly decaying downstream. The azimuthal and wall-normal forces are dominant components that are expected to enhance the Orr and lift-up mechanisms. The resulting optimal response, also peaking at 20 kHz, is essentially the same as the optimal growth disturbance mentioned above except that the initial algebraic growth becomes more pronounced due to external forces.
\par
We have also identified relatively weaker, yet broadband high-frequency optimal growth disturbances that preferentially reside between the boundary layer and the entropy layer on the windward side. They are initially tilted against both the azimuthal and streamwise flow directions, but their orientation with respect to the surface gradually rotates to the opposite direction, suggesting an Orr-like mechanism that mainly amplifies the temperature component of the disturbance.
\par
 Finally, we have drawn a comparison between the optimal growth initiated by the inlet profile and the modal growth obtained separately from local (LST) and multi-dimensional (global) stability analyses (PSE3D). The results from various approaches turn out to differ significantly in terms of the $N$-factor (optimal disturbance reaches the maximum followed by the local and global modal disturbances) as well as the amplitude envelope shape (transition front), highlighting the caution of choosing the stability analysis tool in employing the eN transition prediction method. While local analysis is deemed unable to retrieve the disturbance structures and to tackle the nonlinear interactions of disturbances in a truly 3-D boundary layer, tracing a global mode is of little importance too given the high sensitivity and multiplicity of crossflow eigenvalues \citep{Chen2022}, especially when characterizing the transition mechanism under noisy freestream conditions (e.g. in a conventional wind tunnel). Alternatively, the optimal solution serves as a robust and physically meaningful representation of the crossflow instability, offering a starting point for investigating the nonlinear development and the ensuing breakdown process of crossflow instability disturbances from a global perspective, and moreover, provides an upper bound on the linear growth experienced by disturbances.
 \par
 However, the link between the real freestream disturbances and the optimal profiles is not addressed herein. The recent DNS performed by \cite{Liu2022dns} and \cite{Schuabb2024} for an axisymmetric blunt cone showed that the ``tunnel-like" acoustic disturbance field, which faithfully models the real wind-tunnel noise, could trigger prominent entropy layer disturbances resembling the predictions of optimal growth \citep{Paredes2020}. Their results give confidence that the optimal disturbances are physically realizable. Another important issue that warrants further investigation is figure out whether the present 3-D boundary layer is a low-rank (i.e., with a large gain separation between optimal and suboptimal disturbances) system similar with the 2-D counterpart \citep{Luchini2000}. If so, even though the real disturbances have a small fraction of their energy in the optimal perturbation, one could still expect to observe the response like the optimal one, which would remarkably benefits the transition prediction.
\par
\appendix
\section{Verification of optimal growth calculation}\label{Appendix}
\begin{table}{
\centering
   \begin{tabular}{c c c c c c c c c}
     Case & Model & $Ma$ & $Re_0(X_0^*)$& $Re_1(X_1^*)$& $T_w^*(K)$ & $T_0^*(K)$ & $F (f^*)$ & $\beta$\\
      \hline
      A1 & flat plate &0.001 & 0 & $10000$ & None & 333 & 0 & varying\\
      A2 & flat plate&3 & 500 & 1000 & adiabatic & 333 & 0& varying\\
      A3 & flat plate&6 & 500 & 1000 & $T_w/T_{ad} = 0.8$ & 333 & 0 & varying \\
      B1 & flat plate&6 & 500 & 707  & adiabatic & 430 & 0 & 0.1 \\
      B2 & flat plate&6 & 500 & 707  & adiabatic & 430 &$2\times10^{-4}$ & 0.1 \\
      B3 & flat plate&6 & 500 & 707  & adiabatic & 430 &$3\times10^{-4}$ & 0.1  \\
      C1 & cone at $2^\circ$ AoA&6 & 44 mm & 275 mm & $300$ & 501 &$3.3\times10^{-6}$ (20 kHz) & - \\
      C2 & cone at $2^\circ$ AoA&6 & 44 mm & 275 mm & $300$ & 501 &$8.3\times10^{-6}$ (50 kHz) & -
   \end{tabular}
   \caption{Seven cases of three groups are considered for the verification of the PSE methodology in optimal disturbance calculations. }\label{testcases}}
 \end{table}
The PSE method specifies a single streamwise wave number in order to absorb the main streamwise oscillations of the disturbances into the exponential term, thereby keeping the shape function slowly varying. This casts doubt on whether the PSE method is able to accurately trace the evolution of non-modal disturbances inherently consisting of many modes with varying streamwise wavenumbers \citep{Towne2019}, although PSE and its variant have been widely employed in the non-modal studies \citep[to cite a few]{Levin2003,Guegan2008,Tempelmann2010,Tempelmann2012,Paredes2020}. In order to assess the effects exerted by the PSE method on the optimal growth, we have examined seven cases of three groups as shown in table \ref{testcases}. The cases from the first group come from \cite{Paredes2016ab} who considered the optimal growth for stationary disturbances over a flat-plate boundary layer at several Mach numbers. The second group is to test the ability of the PSE methodology in accurately capturing the non-stationary optimal disturbances by comparing with results of the harmonic linearized Navier-Stokes equations (HLNS) which retain all the elliptic effects. Particularly, the B3 case lies in the second Mack mode regime, and is the most challenging one of the first two groups. Finally, we used DNS to calculate the spatial evolution of the optimal disturbance for two frequencies over the inclined cone at 2 degrees AoA studied in the present work to directly verify the PSE3D's ability in tracing a non-modal disturbance in a 3-D boundary layer.
\par
\begin{figure}
\centering
\includegraphics[width = 0.9\textwidth]{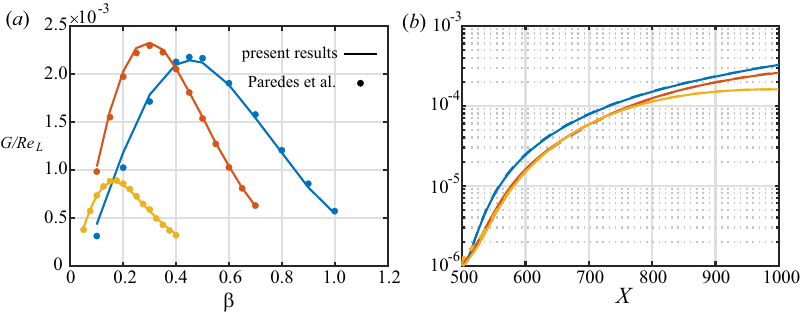}
\caption{($a$) Comparison with optimal gain predictions by \cite{Paredes2016ab} for flat-plate boundary layers at Mach $0.001$ (blue, case A1), 3 (red, case A2) and 6 (yellow, case A3). ($b$) Comparison of the disturbance energy evolution optimized in the range of $X\in(500,1000)$ by the PSE methodology (solid lines) and the HLNS methodology (dashed lines) for three frequencies: $F = 0$ (blue, case B1), $F = 2\times10^{-4}$ (red, case B2) and $F = 3\times10^{-4}$ (yellow, case B3), respectively, at the Mach 6 adiabatic flat-plate boundary layer. } 
\label{PSE-verify}
\end{figure}
\begin{figure}
\centering
\includegraphics[width = 0.9\textwidth]{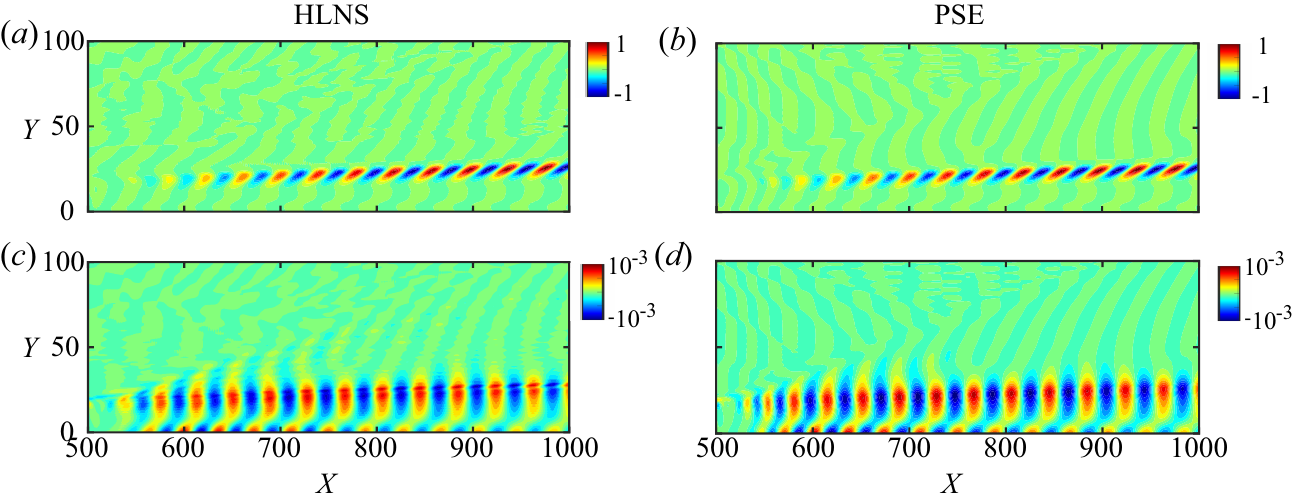}
\caption{Contours of the optimal disturbance normalized by the temperature component for case B3: ($a,c$) HLNS results. ($b,d$) PSE results. First row: temperature fluctuation. Second row: pressure fluctuation.} 
\label{PSE-verify2}
\end{figure}
 The optimal energy gains for the first two groups are displayed in figure \ref{PSE-verify}. It can be seen that the agreement between the present PSE results and the literature results in figure \ref{PSE-verify}($a$), as well as the HLNS results in figure \ref{PSE-verify}($b$) is satisfactory. The fluctuation contours in case B3 are further compared in figure \ref{PSE-verify2}. Again, good agreement between HLNS results and PSE results is observed for the temperature disturbance except for marginal differences in the freestream region. This is reasonable because the freestream disturbances (mainly consisting of acoustic waves) and the fluctuations within the boundary layer have substantially different streamwise wavenumbers (or phase velocities), and might thus not be captured by PSE simultaneously. Compared to the temperature fluctuations, the pressure fluctuations exhibit slightly larger differences, as was also documented by \cite{Towne2019}. This is because the pressure disturbance amplitude is about three order smaller than the temperature disturbance amplitude and small discrepancies in temperature would appear more pronounced in pressure. Since the pressure and freestream disturbances play a secondary role in the non-modal growth process as shown above (see also \cite{Hanifi1996} who noted that the acoustic continuous branches do not contribute to the transient growth of the energy), such discrepancies do not affect the main results of the non-modal analyses.
 \par
\begin{figure}
\centering
\includegraphics[width = 0.9\textwidth]{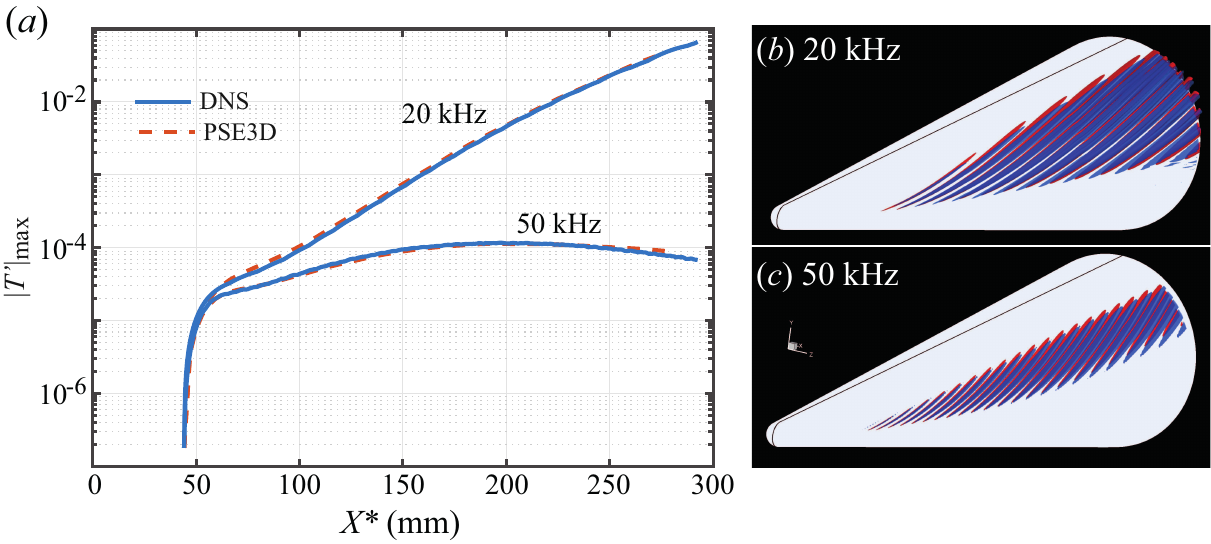}
\caption{Verification of PSE3D results by DNS for the optimal disturbances at 20 kHz and 50 kHz over the inclined cone at 2 degrees AoA. The initial amplitude of DNS is prescribed to a small value to ensure linearity. ($a$) Temperature disturbance amplitude evolution. Temperature disturbance isosurface (real($\hat T$) $=100$, red: DNS, blue: PSE3D) normalized by the initial amplitude for 20 kHz ($b$) and 50 kHz ($c$). } 
\label{AoA2-DNS}
\end{figure}

 At last, we present in figure \ref{AoA2-DNS} the direct comparison between results from DNS and PSE3D for the last group of test cases. The DNS computation domain is the same as the PSE3D. The DNS grid for the lower frequency (most amplified one) is $N_\xi \times N_\eta \times N_\zeta = 1001\times201\times401$ and for the higher one is $N_\xi \times N_\eta \times N_\zeta = 1201\times201\times501$, with slightly more grid points being distributed in the leeward vortex region. Evidently, results from these two approaches match well in terms of the amplitude evolution as well as the spatial structures, thus verifying the PSE3D results.

\bibliographystyle{jfm}
\bibliography{jfm-instructions}

\begin{thebibliography}{56}
\expandafter\ifx\csname natexlab\endcsname\relax\def\natexlab#1{#1}\fi
\def\au#1{#1} \def\ed#1{#1} \def\yr#1{#1}\def\at#1{#1}\def\jt#1{\textit{#1}}
  \def\bt#1{#1}\def\bvol#1{\textbf{#1}} \def\vol#1{#1} \def\pg#1{#1}
  \def\publ#1{#1}\def\arxiv#1{#1}\def\org#1{#1}\def\st#1{\textit{#1}}

\bibitem[{\AA}kervik {\em et~al.\/}(2008){\AA}kervik, U, Gallarire \&
  Henningson]{Akervik2008}
{\sc \au{{\AA}kervik, E.}, \au{U, Ehrenstein}, \au{Gallarire, F.} \&
  \au{Henningson, D.~S.}} \yr{2008}  \at{Global two-dimensional stability
  measures of the flat plate boundary layer flow}.  \jt{Eur. J. Mech B}
  \bvol{27},  \pg{501--513}.

\bibitem[Alizard \& Robinet(2007)]{Alizard2007}
{\sc \au{Alizard, F.} \& \au{Robinet, J.-C.}} \yr{2007}  \at{Spatially
  convective global modes in a boundary layer}.  \jt{Phys. Fluids}
  \bvol{19}~(114105).

\bibitem[Araya {\em et~al.\/}(2023)Araya, Bitter, Wheaton, Kamal, Colonius,
  Knutson, Johnson, Nichols, Candler, Russo \& Brehm]{Araya2023}
{\sc \au{Araya, D.}, \au{Bitter, N.}, \au{Wheaton, B.}, \au{Kamal, O.},
  \au{Colonius, T.}, \au{Knutson, A.}, \au{Johnson, H.}, \au{Nichols, J.},
  \au{Candler, G.}, \au{Russo, V.} \& \au{Brehm, C.}} \yr{2023}  \at{Linear
  analysis of boundary-layer instabilities on a finned-cone at \rm{Mach} 6}.
  \jt{arXiv} ~(2303.10747).

\bibitem[Bagheri {\em et~al.\/}(2009)Bagheri, {\AA}kervik, Brandt \&
  Henningson]{Bagheri2009}
{\sc \au{Bagheri, S.}, \au{{\AA}kervik, E.}, \au{Brandt, L.} \& \au{Henningson,
  D.~S.}} \yr{2009}  \at{Matrix-free methods for the stability and control of
  boundary layers}.  \jt{AIAA J.}  \bvol{47}~(5),  \pg{1057--1068}.

\bibitem[Bippes(1999)]{Bippes1999}
{\sc \au{Bippes, H.}} \yr{1999}  \at{Basic experiments on transition in
  three-dimensional boundary layers dominated by crossflow instability}.
  \jt{Prog. Aerosp. Sci}  \bvol{35}~(363-412).

\bibitem[Bugeat {\em et~al.\/}(2019)Bugeat, Chassing, Robinet \&
  Sagaut]{Bugeat2019}
{\sc \au{Bugeat, B.}, \au{Chassing, J.}, \au{Robinet, J.} \& \au{Sagaut, P.}}
  \yr{2019}  \at{\rm{3D} global optimal forcing and response of the supersonic
  boundary layer}.  \jt{J. Comput. Phys.}  \bvol{398}~(108888).

\bibitem[Butler \& Farrell(1992)]{Butler1992}
{\sc \au{Butler, K.~M.} \& \au{Farrell, B.~F.}} \yr{1992}
  \at{Three-dimensional optimal perturbations in viscous shear flow}.
  \jt{Phys. Fluids}  \bvol{4},  \pg{1637--1650}.

\bibitem[Chen {\em et~al.\/}(2022)Chen, Dong, Tu, Yuan \& Chen]{Chen2022}
{\sc \au{Chen, X.}, \au{Dong, S.}, \au{Tu, G.}, \au{Yuan, X.} \& \au{Chen, J.}}
  \yr{2022}  \at{Boundary layer transition and linear modal instabilities of
  hypersonic flow over a lifting body}.  \jt{J.~Fluid Mech.}  \bvol{938}~(A8).

\bibitem[Chen {\em et~al.\/}(2023{\natexlab{{\em a\/}}})Chen, Dong, Tu, Yuan \&
  Chen]{Chen2023PRF}
{\sc \au{Chen, X.}, \au{Dong, S.}, \au{Tu, G.}, \au{Yuan, X.} \& \au{Chen, J.}}
  \yr{2023{\natexlab{{\em a\/}}}}  \at{Global stability analyses of \rm{Mack}
  mode on the windward face of a hypersonic yawed cone}.  \jt{Phys. Rev.
  Fluids}  \bvol{8}~(033903).

\bibitem[Chen {\em et~al.\/}(2023{\natexlab{{\em b\/}}})Chen, Tu, Wan, Yuan,
  Chen \& Chen]{Chen2023b}
{\sc \au{Chen, X.}, \au{Tu, G.}, \au{Wan, B.}, \au{Yuan, X.}, \au{Chen, J.} \&
  \au{Chen, J.}} \yr{2023{\natexlab{{\em b\/}}}}  \at{The leeward vortex
  transition over hypersonic yawed cones by \rm{eN} method based on global
  stability theory (in \rm{Chinese})}.  \jt{Acta Aerodynamica Sinica}
  \bvol{41}~(X).

\bibitem[Chomaz(2005)]{Chomaz2005}
{\sc \au{Chomaz, J.}} \yr{2005}  \at{Global instabilities in spatially
  developing flows: non-normality and nonlinearity}.  \jt{Annu. Rev. Fluid
  Mech.}  \bvol{37},  \pg{357--392}.

\bibitem[Chu(1965)]{Chu1965}
{\sc \au{Chu, B.~T.}} \yr{1965}  \at{On the energy transfer to small
  disturbances in fluid flow (part 1)}.  \jt{Acta Mechanica}  \bvol{1}~(3),
  \pg{215--234}.

\bibitem[Cook {\em et~al.\/}(2020)Cook, Knutson \& Nichols]{Cook2020}
{\sc \au{Cook, D.}, \au{Knutson, A.} \& \au{Nichols, J.}} \yr{2020}  \at{Matrix
  methods for input-output analysis of \rm{2D} and \rm{3D} hypersonic flows}.
  \jt{AIAA Paper 2020-1820} .

\bibitem[Corbett \& Bottaro(2001)]{Corbett2001}
{\sc \au{Corbett, P.} \& \au{Bottaro, A.}} \yr{2001}  \at{Optimal linear growth
  in swept boundary layers}.  \jt{J. ~Fluid Mech.}  \bvol{435},  \pg{1--23}.

\bibitem[Cossu \& Chomaz(1997)]{Cossu1997}
{\sc \au{Cossu, C.} \& \au{Chomaz, J.~M.}} \yr{1997}  \at{Global measures of
  local convective instabilities}.  \jt{Phys. Rev. Lett.}  \bvol{78},
  \pg{4387--4390}.

\bibitem[Ehrenstein \& Gallaire(2005)]{Ehrenstein2005}
{\sc \au{Ehrenstein, U.} \& \au{Gallaire, F.}} \yr{2005}  \at{On
  two-dimensional temporal modes in spatially evolving open flows: the
  flat-plate boundary layer}.  \jt{J.~Fluid Mech.}  \bvol{536},  \pg{209--218}.

\bibitem[Farrell \& Ioannou(1993)]{Farrell1993}
{\sc \au{Farrell, B.~F.} \& \au{Ioannou, P.~J.}} \yr{1993}  \at{Optimal
  excitation of three-dimensional perturbations in viscous constant shear
  flow}.  \jt{Phys. Fluids}  \bvol{5}.

\bibitem[Giannetti \& Luchini(2007)]{Giannetti2007}
{\sc \au{Giannetti, F.} \& \au{Luchini, P.}} \yr{2007}  \at{Structural
  sensitivity of the first instability of the cylinder wake}.  \jt{J.~Fluid
  Mech.}  \bvol{581}~(1),  \pg{167--197}.

\bibitem[Guegan {\em et~al.\/}(2008)Guegan, Schmid \& Huerre]{Guegan2008}
{\sc \au{Guegan, A.}, \au{Schmid, P.~J.} \& \au{Huerre, P.}} \yr{2008}
  \at{Spatial optimal disturbances in swept attachment-line boundary layers}.
  \jt{J.~Fluid Mech.}  \bvol{603},  \pg{179--188}.

\bibitem[Hanifi {\em et~al.\/}(1996)Hanifi, Schmid \& Henningson]{Hanifi1996}
{\sc \au{Hanifi, A.}, \au{Schmid, P.~J.} \& \au{Henningson, D.~S.}} \yr{1996}
  \at{Transient growth in compressible boundary layer flow}.  \jt{Phys. Fluids}
   \bvol{8}~(3),  \pg{826--837}.

\bibitem[Jovanovi\'{c} \& Bamieh(2005)]{Jovanovic2005}
{\sc \au{Jovanovi\'{c}, M.~R.} \& \au{Bamieh, B.}} \yr{2005}  \at{Componentwise
  energy amplification in channel flows}.  \jt{J. Fluid Mech.}  \bvol{534},
  \pg{145--183}.

\bibitem[Kocian {\em et~al.\/}(2019)Kocian, Moyes, Reed, Craig, Saric,
  Schneider \& Edelman]{Kocian2019}
{\sc \au{Kocian, T.~S.}, \au{Moyes, A.~J.}, \au{Reed, H.~L.}, \au{Craig,
  S.~A.}, \au{Saric, W.~S.}, \au{Schneider, S.~P.} \& \au{Edelman, J.~B.}}
  \yr{2019}  \at{Hypersonic crossflow instability}.  \jt{J. Spacecr. Rockets}
  \bvol{56}~(2),  \pg{432--446}.

\bibitem[Laible \& Fasel(2016)]{Laible2016}
{\sc \au{Laible, A.~C.} \& \au{Fasel, H.~F.}} \yr{2016}  \at{Continuously
  forced transient growth in oblique breakdown for supersonic boundary layers}.
   \jt{J.~Fluid Mech.}  \bvol{804},  \pg{323--350}.

\bibitem[Lakebrink {\em et~al.\/}(2017)Lakebrink, Paredes \&
  Borg]{Lakebrink2017}
{\sc \au{Lakebrink, M.~T.}, \au{Paredes, P.} \& \au{Borg, M.~P.}} \yr{2017}
  \at{Toward robust prediction of crossflow-wave instability in hypersonic
  boundary layers}.  \jt{Comput. Fluids}  \bvol{144},  \pg{1--9}.

\bibitem[Levin \& Henningson(2003)]{Levin2003}
{\sc \au{Levin, O.} \& \au{Henningson, D.~S.}} \yr{2003}  \at{Exponential vs
  algebraic growth and transition prediction in boundary layer flow}.
  \jt{Flow, Turbul. Combust.} ~(70),  \pg{183--210}.

\bibitem[Liu {\em et~al.\/}(2022{\natexlab{{\em a\/}}})Liu, Wan, Yuan, Zhang,
  Chen \& Chen]{Liu2022}
{\sc \au{Liu, S.}, \au{Wan, B.}, \au{Yuan, X.}, \au{Zhang, L.}, \au{Chen, J.}
  \& \au{Chen, X.}} \yr{2022{\natexlab{{\em a\/}}}}  \at{Linear modal global
  instabilities of hypersonic flow over an inclined cone}.  \jt{Phys. Fluids}
  \bvol{34}~(074108).

\bibitem[Liu {\em et~al.\/}(2022{\natexlab{{\em b\/}}})Liu, Schuabb, Duan,
  Paredes \& Choudhari]{Liu2022dns}
{\sc \au{Liu, Y.}, \au{Schuabb, M.}, \au{Duan, L.}, \au{Paredes, P.} \&
  \au{Choudhari, M.~M.}} \yr{2022{\natexlab{{\em b\/}}}}  \at{Interaction of a
  tunnel-like acoustic disturbance field with a blunt cone boundary layer at
  \rm{Mach} 8}.  \jt{AIAA Paper 2022-3250} .

\bibitem[Luchini(2000)]{Luchini2000}
{\sc \au{Luchini, P.}} \yr{2000}  \at{Reynolds-number-independent instability
  of the boundary layer over a flat surface: optimal perturbations}.
  \jt{J.~Fluid Mech.}  \bvol{404},  \pg{289--309}.

\bibitem[Nastro {\em et~al.\/}(2020)Nastro, Fontane \& Joly]{Nastro2020}
{\sc \au{Nastro, G.}, \au{Fontane, J.} \& \au{Joly, L.}} \yr{2020}  \at{Optimal
  perturbations in viscous round jets subject to \rm{Kelvin-Helmholtz}
  instability}.  \jt{J. ~Fluid Mech.}  \bvol{900}~(A13).

\bibitem[Nichols \& Lele(2011)]{Nichols2011}
{\sc \au{Nichols, J.~W.} \& \au{Lele, S.~K.}} \yr{2011}  \at{Global modes and
  transient response of a cold supersonic jet}.  \jt{J.~Fluid Mech.}
  \bvol{669},  \pg{225--241}.

\bibitem[Paredes {\em et~al.\/}(2020)Paredes, Choudhari \& Li]{Paredes2020}
{\sc \au{Paredes, P.}, \au{Choudhari, M.} \& \au{Li, F.}} \yr{2020}
  \at{Mechanism for frustum transition over blunt cones at hypersonic speeds}.
  \jt{J.~Fluid Mech.}  \bvol{894}~(A22).

\bibitem[Paredes {\em et~al.\/}(2016{\natexlab{{\em a\/}}})Paredes, Choudhari,
  Li \& Chang]{Paredes2016ab}
{\sc \au{Paredes, P.}, \au{Choudhari, M.}, \au{Li, F.} \& \au{Chang, C-L}}
  \yr{2016{\natexlab{{\em a\/}}}}  \at{Optimal growth in hypersonic boundary
  layers}.  \jt{AIAA J.}  \bvol{54}~(10),  \pg{3050--3061}.

\bibitem[Paredes {\em et~al.\/}(2019)Paredes, Choudhari, Li, Jewell \&
  Kimmel]{Paredes2019}
{\sc \au{Paredes, P.}, \au{Choudhari, M.}, \au{Li, F.}, \au{Jewell, J.~S.} \&
  \au{Kimmel, R.~L.}} \yr{2019}  \at{Nonmodal growth of traveling waves on
  blunt cones at hypersonic speeds}.  \jt{AIAA paper 2019-0876} .

\bibitem[Paredes {\em et~al.\/}(2016{\natexlab{{\em b\/}}})Paredes, Gosse,
  Theofilis \& Kimmel]{Paredes2016a}
{\sc \au{Paredes, P.}, \au{Gosse, R.}, \au{Theofilis, V.} \& \au{Kimmel, R.}}
  \yr{2016{\natexlab{{\em b\/}}}}  \at{Linear modal instabilities of hypersonic
  flow over an elliptic cone}.  \jt{J. Fluid Mech.}  \bvol{804},
  \pg{442--466}.

\bibitem[Paredes {\em et~al.\/}(2023)Paredes, Scholten, Choudhari \&
  Li]{Paredes2023JSR}
{\sc \au{Paredes, P.}, \au{Scholten, A.}, \au{Choudhari, M.} \& \au{Li, F.}}
  \yr{2023}  \at{Modal instabilities over blunted cones at angle of attack in
  hypersonic flow}.  \jt{J. Spacecraft Rockets}  \bvol{60}~(4).

\bibitem[Paredes {\em et~al.\/}(2022)Paredes, Scholten, Choudhari, Li, Price \&
  Jewell]{Paredes2022}
{\sc \au{Paredes, P.}, \au{Scholten, A.}, \au{Choudhari, M.}, \au{Li, F.},
  \au{Price, B.} \& \au{Jewell, J.}} \yr{2022}  \at{Combined bluntness and
  roughness effects on cones at hypersonic speeds}.  \jt{AIAA 2022-3340} .

\bibitem[Quintanilha {\em et~al.\/}(2022)Quintanilha, Paredes \&
  Hanifi]{Quintanilha2022}
{\sc \au{Quintanilha, H.}, \au{Paredes, P.} \& \au{Hanifi, A.}} \yr{2022}
  \at{Transient growth analysis of hypersonic flow over an elliptic cone}.
  \jt{J. Fluid Mech.}  \bvol{935}~(A40).

\bibitem[Reddy {\em et~al.\/}(1993)Reddy, Schmid \& Henningson]{Reddy1993}
{\sc \au{Reddy, S.}, \au{Schmid, P.~J.} \& \au{Henningson, D.~S.}} \yr{1993}
  \at{Pseudospectra of the \rm{Orr-Sommerfeld} operator}.  \jt{SIAM J. Appl.
  Math.}  \bvol{53}~(1),  \pg{15--47}.

\bibitem[Reed \& Saric(1989)]{Reed1989}
{\sc \au{Reed, H.~L.} \& \au{Saric, W.~S.}} \yr{1989}  \at{Stability of
  three-dimensional boundary layers}.  \jt{Ann. Rev. Fluid Mech.}  \bvol{21},
  \pg{235--284}.

\bibitem[Reshotko \& Tumin(2000)]{Reshotko2000}
{\sc \au{Reshotko, E.} \& \au{Tumin, A.}} \yr{2000}  \at{The blunt body paradox
  - a case for transient growth}.  \jt{Laminar-Turbulence Transition, IUTAM
  Symposium, Sedona, edited by H. F. Fasel and W. S. Saric, Springer-Verlag,
  Berlin}  \pg{pp. 403--408}.

\bibitem[Schmid(2007)]{Schmid2007}
{\sc \au{Schmid, P.~J.}} \yr{2007}  \at{Nonmodal stability theory}.  \jt{Annu.
  Rev. Fluid Mech.}  \bvol{39},  \pg{129--162}.

\bibitem[Schmid \& Brandt(2014)]{Schmid2014}
{\sc \au{Schmid, P.~J.} \& \au{Brandt, L.}} \yr{2014}  \at{Analysis of fluid
  systems: stability, receptivity, sensitivity: lecture notes from the
  \rm{FLOW-NORDITA} summer school on advanced instability methods for complex
  flows}.  \jt{Appl. Mech. Rev.}  \bvol{66}~(024803).

\bibitem[Schmid \& Henningson(2001)]{Schmid2001}
{\sc \au{Schmid, P.~J.} \& \au{Henningson, D.~S.}} \yr{2001} {\em Stability and
  Transition in Shear Flows\/}.  \publ{New York: Springer-Verlag}.

\bibitem[Schuabb {\em et~al.\/}(2024)Schuabb, Duan, Scholten, Paredes \&
  Choudhari]{Schuabb2024}
{\sc \au{Schuabb, M.}, \au{Duan, L.}, \au{Scholten, A.}, \au{Paredes, P.} \&
  \au{Choudhari, M.~M.}} \yr{2024}  \at{Hypersonic boundary-layer transition
  over a blunt circular cone in a \rm{Mach} 8 digital wind tunnel}.  \jt{AIAA
  Paper 2024-2181} .

\bibitem[Skene {\em et~al.\/}(2022)Skene, Yeh, Schmid \& Taira]{Skene2022}
{\sc \au{Skene, C.~S.}, \au{Yeh, C.-A.}, \au{Schmid, P.~J.} \& \au{Taira, K.}}
  \yr{2022}  \at{Sparsifying the resolvent forcing mode via gradient-based
  optimisation}.  \jt{J.~Fluid Mech.}  \bvol{944}~(A52).

\bibitem[Song {\em et~al.\/}(2020)Song, Zhao \& Huang]{Song2020}
{\sc \au{Song, R.}, \au{Zhao, L.} \& \au{Huang, Z.}} \yr{2020}  \at{Improvement
  of the parabolized stability equation to predict the linear evolution of
  disturbances in three-dimensional boundary layers based on ray tracing
  theory}.  \jt{Phys. Rev. Fluids}  \bvol{5}~(033901).

\bibitem[Tempelmann {\em et~al.\/}(2010)Tempelmann, Hanifi \&
  Henningson]{Tempelmann2010}
{\sc \au{Tempelmann, D.}, \au{Hanifi, A.} \& \au{Henningson, D.~S.}} \yr{2010}
  \at{Spatial optimal growth in three-dimensional boundary layers}.  \jt{J.
  Fluid Mech.}  \bvol{646},  \pg{5--37}.

\bibitem[Tempelmann {\em et~al.\/}(2012)Tempelmann, Hanifi \&
  Henningson]{Tempelmann2012}
{\sc \au{Tempelmann, D.}, \au{Hanifi, A.} \& \au{Henningson, D.~S.}} \yr{2012}
  \at{Spatial optimal growth in three-dimensional compressible boundary
  layers}.  \jt{J. Fluid Mech.}  \bvol{704},  \pg{251--279}.

\bibitem[Towne {\em et~al.\/}(2019)Towne, Rigas \& Colonius]{Towne2019}
{\sc \au{Towne, A.}, \au{Rigas, G.} \& \au{Colonius, T.}} \yr{2019}  \at{A
  critical assessment of the parabolized stability equations}.  \jt{Theor.
  Comput. Fluid Dyn.}  \bvol{33},  \pg{359--382}.

\bibitem[Trefethen \& Embree(2005)]{Trefethen2005}
{\sc \au{Trefethen, L} \& \au{Embree, M}} \yr{2005} {\em Spectra and
  Pseudospectra\/}.  \publ{Princeton: Princeton University Press}.

\bibitem[Tumin \& Reshotko(2001)]{Tumin2001}
{\sc \au{Tumin, A.} \& \au{Reshotko, E.}} \yr{2001}  \at{Spatial theory of
  optimal disturbances in boundary layers}.  \jt{Phys. Fluids}
  \bvol{13}~(2097).

\bibitem[Tumin \& Reshotko(2003)]{Tumin2003}
{\sc \au{Tumin, A.} \& \au{Reshotko, E.}} \yr{2003}  \at{Optimal disturbances
  in compressible boundary layers}.  \jt{AIAA J.}  \bvol{41}~(12),
  \pg{2357--2363}.

\bibitem[Vigneron {\em et~al.\/}(1978)Vigneron, Rakich \&
  Tannehill]{Vigneron1978}
{\sc \au{Vigneron, Y.}, \au{Rakich, J.} \& \au{Tannehill, J.}} \yr{1978}
  \at{Calculation of supersonic viscous flow over delta wings with sharp
  supersonic leading edges}.  \jt{AIAA Paper 1978-1137} .

\bibitem[Wang {\em et~al.\/}(2023)Wang, Xiang, Dong, Yuan, Chen \&
  Chen]{Wang2023}
{\sc \au{Wang, Q.}, \au{Xiang, X.}, \au{Dong, S.}, \au{Yuan, X.}, \au{Chen, J.}
  \& \au{Chen, X.}} \yr{2023}  \at{Wall temperature effects on the hypersonic
  boundary-layer transition over an inclined blunt cone}.  \jt{Phys. Fluids}
  \bvol{35}~(024107).

\bibitem[Wassermann \& Kloker(2003)]{Wassermann2003}
{\sc \au{Wassermann, P.} \& \au{Kloker, M.}} \yr{2003}  \at{Transition
  mechanisms induced by travelling crossflow vortices in a three-dimensional
  boundary layer}.  \jt{J.~Fluid Mech.}  \bvol{483},  \pg{67--89}.

\bibitem[Zhao(2017)]{Zhao2017}
{\sc \au{Zhao, L.}} \yr{2017}  \at{Flow stability and wave propagation in
  hypersonic three-dimensional boundary layer}.  \jt{PhD thesis, Tianjin
  University} .

\end{thebibliography}

\end{document}